\documentclass[twocolumn,english,aps,superscriptaddress,pra]{revtex4-1}

\usepackage{array}
\usepackage[latin9]{inputenc}
\usepackage{amsmath}
\usepackage{amssymb}
\usepackage{graphicx}
\usepackage{babel}
\usepackage{mathrsfs}
\usepackage{amsfonts}
\usepackage{epstopdf}
\usepackage{multirow}
\usepackage{color}
\usepackage{natbib}

\begin{document}
\title{Topological Quantum Critical Points in Strong Coupling limits:\\
 Global Symmetries and Strongly Interacting Majorana Fermions}

\author{Fei Zhou}

\address{Department of Physics and Astronomy, University of British Columbia, 6224 Agricultural Road, Vancouver, BC, V6T 1Z1, Canada}

\begin{abstract}
In this article, we discuss strong coupling limits of topological quantum critical points (TQCPs) where quantum phase transitions between two topological distinct superconducting states take place.
We illustrate that  while superconducting phases on both sides of TQCPs spontaneously break same symmetries,
universality classes of critical states can be identified only when global symmetries in topological states are further specified.
In dimensions $d=2,3$, we find that continuous $(d+1)$th order transitions at weakly interacting TQCPs that were pointed out previously in the presence of emergent Lorentz symmetry can be terminated by
strongly interacting fixed points of majorana fields. For $2d$ time reversal symmetry breaking TQCPs,
termination points are supersymmetric with ${\mathcal N}=4N_f={1}$ (where $N_f$ is the number of four-component Dirac fermions and ${\mathcal N}$ is the number of two-component real fermions) beyond which transitions are discontinuous first order ones.
For $2d$ time reversal symmetric TQCPs without other global symmetries, termination points of $(d+1)$th order continuous transition lines 
are generically conformal invariant without supersymmetry.
Beyond these strong coupling fixed points, there are first-order discontinuous transitions as far as the protecting symmetry is not spontaneously broken but
no direct transitions if the protecting symmetry is spontaneously broken in the presence of strong interactions. 
In this case, termination points can become supersymmetric with ${\mathcal N}=4N_f= 2$ only when interacting majorana fermions further display an emergent global $U(1)$ symmetry.  
In $3d$, strong coupling termination points can be further effectively represented by new emergent gapless real bosons weakly coupled with free gapless majorana fermions.
However, in $1d$, time reversal symmetric $(d+1)$th continuous transition lines of TQCPs are terminated by simple free majorana fermion fixed points.
\end{abstract}

\date{\today}
\maketitle

\section{Introduction}

Quantum phase transitions between topological states with distinctly different global topologies\cite{Schnyder08,Kitaev09} have been fascinating.
There are at least two subclasses of such phenomena known to us so far. One are transitions that are entirely driven and defined by a change of topologies or topological invariants and otherwise do not
exist in their non-topological counter-parts. A familiar example of
these unique transitions is the one due to increasing interactions in a topological superconductor. 
Because there can be a change of global topology when interactions allowed by symmetries increase, even though states are ordered in the same way locally,
there shall be a resultant bulk transition beyond the standard Landau paradigm of order-disorder transitions.
These transitions usually turn out to be of
relatively higher order. This is in stark contrast to
what happens in a non-topological s-wave superconductor
when one varies the same parameters. There, if one
increases the interaction strength from weak to strong,
there are no transitions in conventional s-wave superconductors, because an $s$-wave superconductor can only be topologically trivial. 
So the very existence of such transitions in topological superconductors entirely and crucially relies on the notion of global topology
of underlying superconductors and topological distinction of different states
\cite{Schnyder08,Kitaev09,Qi10,Qi08,Gurarie11,Volovik09,Wang12,Volovik88,Volovik03,Read00,Qi11,Bernevig13,Hasan10,Fu08}.
In other words, global topologies and their characterization add a new dimension to the
parameter space along which phase transitions can occur. This new dimension is beyond the standard Landau
paradigm of order-disorder transitions.

The second subclass are transitions which can also occur in more conventional non-topological states but for transitions in topological states, changes of topologies are further involved.
In all the cases we have examined, changes of topologies result in distinct universality classes that are different from their counter parts in non-topological systems.
For instance, when an external Zeeman field is applied, an s-wave superconductor can have a transition to an FFLO state that further breaks translational symmetry in additional to U(1) symmetry.
In topological superconductors or superfluids, an external Zeeman field can drive a time-reversal-invariant topological state at zero field into a nodal phase with topological nodal points in its spectrum. 
The resultant state is a topological state with perfect translational symmetry.


These TQCPs do not appear in
the Landau paradigm of order-disorder phase transitions
because the states on both sides of TQCPs and
TQCPs themselves either all spontaneously break same symmetries or all have the same symmetries. Therefore, TQCPs 
do not involve condensation of new bosonic quantum
fields or particles. It is exactly the global topology and
topological distinctions between different states that result in such a new class of quantum critical points. {\em And topological phase transitions associated with these quantum critical points only exist at zero temperature. }
 For this reason, we have referred them as topological quantum critical points, TQCPs.

As changes of topological invariants are discrete rather than continuous, one might wonder how it is possible to describe 
TQCPs in terms of  continuum quantum fields.
Such a possibility of a continuum quantum
field theory representation of topological quantum criticality can be most conveniently understood via employing a standard adiabatic theorem. 

Consider a fully gapped fermonic topological state protected by global symmetries. One anticipates that it can maintain its
topological distinction under small Hamiltonian deformations within its symmetry class if the gap remains open. A change in global
topologies typically implies closing the gap. If the gap indeed closes so to change global topologies or transitions are continuous, then there shall be coalescing of gapped fermions into a ground state
near TQCPs. More importantly, in superconductors and fermonic superfluids, elementary emergent particles that are relevant to generic TQCPs are usually fermonic, not bosonic fields. In
fact, they shall be real Majorana fermions due to the emergent
charge conjugation symmetry at U(1) symmetry breaking
TQCPs. So generically, if these topological quantum phase
transitions occur, we anticipate coalescing of real fermions into the
ground state without condensation of new bosonic quantum fields
or spontaneously breaking additional symmetries. This
aspect is fundamentally different from the Landau paradigm of order-disorder transitions.
On the other hand, the above observation does explicitly suggest effective quantum fields
for coalesce dynamics of gapped particles and
topological quantum criticality in superconductors.

One can further extend above arguments to situations where topological states are gapless nodal phases.
Gapless topological states are characterized by topological invariants in an embedded subspace instead of a full physical space\cite{Sato17,Wen02,Kobayashi14,Zhao16,Meng12,Cho12,Sato06,Schnyder11,Beri10}.
Compared with fully gapped topological states, this aspect of gapless phases makes a big difference in constructing quantum field theory representations at TQCPs.
At TQCPs, topological charges embedded in different subspaces are created (or annihilated) and this is achieved again via fermion coalescing.   
For instance, charges associated with nodal points can be created in pairs at TQCPs when a Zeeman field is applied to a fully gapped topological state.   
This happens when the gap closes at a point or a few discrete points and real fermions coalesce along a particular direction of Zeeman or spin-exchange fields.
And different topologies of quantum nodal phases
naturally require different quantum field theory representations and hence result in different universality classes.
For instance, a nodal point topology demands very different quantum Lifshitz majorana fields than a nodal line
topology would demand.

Perhaps the most remarkable consequence of gap closing
in topological states is proliferation of surface states
into an interior of topological matter. This is essential
at TQCPs so that a state can topologically reconstruct in the bulk and boundary simultaneously across a TQCP. 
Physically, these bulk transitions are always accompanied or even heralded by surface
quantum criticality, a very distinct aspect of topological
quantum criticality. 
A continuum quantum field representation shall properly reflect such a {\em bulk-boundary} correspondence at generic TQCPs.

These two particular aspects, one
reflecting bulk topologies and the other more on their
consequences at boundaries, are both absent in other gap
closing phenomena in conventional non-topological transitions or in the standard Landau paradigm. Actually these two aspects are what makes 
TQCPs outstanding. Therefore, any appropriate quantum
field theory representations for topological TQCPs have to
be in a class of field theories with boundary states reflecting a corresponding change of topologies
across TQCPs. This is one of the guiding principles in explicit constructions, apart from more
standard symmetry considerations. The unique role of
changes of topologies in these zero temperature transitions therefore has to be encoded in these effective fields.

In this article, we will focus on the roles of global symmetries on dynamics of effective fields and hence TQCPs.
In order-disorder phase transitions in the standard Landau paradigm, symmetries of Hamiltonians and spontaneous breaking of them play a paramount role in constructing
effective quantum field theories. At TQCPs that are not driven by further spontaneous breaking of symmetries, one might wonder what is the role of symmetries then.

Although TQCPs considered below are entirely driven by changes of topologies, not by condensation of bosonic particles, global symmetries {\em do} still play a paramount role.
That is perhaps not surprising from the point of view of gapped topological states that are symmetry protected anyway\cite{Chen10,Chen12,Chen13,Kane05,Bernevig06b,Fu07, Moore07}. 
Since topologies are classified in terms of global symmetries and 
topological states are symmetry protected, TQCPs that terminate
those phases shall rely on protecting symmetries as well. Indeed, dynamics of interacting fermion fields emerging at TQCPs crucially depend on at least two classes of fundamental global symmetries.

The first important symmetry is the continuous $U(1)$ symmetry associated with charge conservation. 
Exactly at standard quantum critical points in the Landau paradigm, continuous symmetries such as $U(1)$ symmetry is unbroken; they are only broken in one of 
phases next to quantum critical points. TQCPs are unique in a sense that such continuous global symmetries can be broken in the entire vicinity of a TQCP and the entire quantum critical regime
supports the same local order. 

If TQCPs are $U(1)$ symmetric, complex fermion fields that can exhibit global $U(1)$ symmetries shall emerge and form a faithful representation.
If $U(1)$ symmetry itself is spontaneously broken at TQCPs and in all quantum phases in adjacent to them, one needs to adopt majorana fermion representations that are suitable.
Most of discussions in this article will be on $U(1)$ symmetry broken superconductors or superfluids, although many discussions can be extended to $U(1)$ symmetric TQCPs.
Detailed dynamics of fermion fields can further depend on other continuous symmetries. For instance, additional continuous symmetry breaking such rotational ones at TQCPs and in their proximity can 
result in distinct classes of scale symmetries that we can associate with generalized fermionic Lifshitz fields\cite{Yang21,Lifshitz41}.

The second class are discrete global symmetries such as time reversal symmetry and/or emergent charge conjugation symmetry etc.  These symmetries have played a paramount role in 
defining topological winding numbers and classifying topological states so naturally they also play an important role at TQCPs. In Section IV, we will focus on such aspect at TQCPs.
In general, a larger global symmetry requires a bigger fermion representation. Physically, it naturally leads to more emergent fermionic fields at TQCPs.
The role of these global symmetries is one of the main focuses of this article. We illustrate dominating effects of global symmetries in both weakly and strongly interacting limits,
with an emphasis on the interplay between symmetries and strong interactions.

In Sec II, we will introduce general phenomenologies of TQCPs.
This part of discussions are pedagogical. We summarize previous discussions on topological state classifications via Green's functions of interacting fermions and then further extend and apply to near TQCPs. 
In Sec III, we identify a few main issues in strong coupling limits that are focused on in this article.
In Sec. IV, we discuss the main role of global symmetries on dynamics at TQCPs and illustrate an intimate relation between global symmetries and numbers of emergent low energy majorana degrees of freedom.
In Sec V, VI, we will elaborate on dominating effects of global symmetries on TQCPS in various strong coupling limits. 
In Section V, we focus on TQCPs in topological superconductors with no global symmetries and illustrate that there shall always be a transition between two topologically distinct states
even in strong coupling limits. However, continuous $(d+1)$ order transition lines pointed out in a previous study in weakly coupling limits\cite{Yang19} shall be terminated by a strong coupling supersymmetric conformal fixed point
beyond which transitions become first order ones.  
In Sec VI, we study TQCPs with a single global symmetry (time reversal symmetry) but no other global symmetries.
We emphasize peculiar aspects of an emergent $U(1)$ symmetry when interactions are local and various mass generation mechanisms that can spontaneously 
break the protecting time reversal symmetry. We found that when time reversal global symmetry is not spontaneously broken by strong interactions, $(d+1)$th order transition lines are again terminated by a conformal fixed point beyond which
first order transitions occur. However, when the protecting time reversal symmetry is broken spontaneously by strong interactions, the continuous $(d+1)$th order transition line is  simply terminated by either supersymmetric or 
non-supersymmetric conformal field fixed points. Beyond those points, two topological states can be deformed into each other in the absence of protecting symmetries.
 In Sec. VII, we further discuss some open questions on TQCPs in topological superconductors and possible future directions.
Sec.IX concludes our article.

\section{General Phenomenology of TQCPs}

\subsection{Quantum Field Representations for TQCPs}

A TQCP, if exists, connects two topologically distinct states.
To faithfully reflect such an aspect of TQCPs in a quantum field representation, in additional to standard symmetry considerations in constructing effective theories,
we can further require Green's functions of interacting quantum fields being a representation of a homotopy group that defines topologies or topological invariants (see Fig.1). 
Similar ideas had been previously introduced to understand topological quantum field representations for topological insulators and Witten effects and axion dynamics near boundaries\cite{Qi08,Qi11}.
One can also naturally apply it, as a general principle, to TQCPs
for the main purpose to understand bulk dynamics near a TQCP and later on the bulk-boundary correspondence.

We first consider transitions between two fully gapped superconducting states, one topological but the other is trivial.
Let us define a tuning parameter that drives such a transition as $m=\mu$ which depends on microscopic parameters; TQCPs correspond to $\mu=\mu_c$.
To apply to $(d+1)$ dimension space-time of bulk states near TQCPs, we can further extend the space to $(d+2)$ dimension by inclusion of an additional dimension of the general tuning parameter $\mu$ along which a TQCP occurs.
Green's functions defined in $(d+2)$ dimension, $G(i\Omega, {\bf k};\mu)$ are $M \times M$ complex matrices where $M$ are numbers of bands under considerations. 
These functions shall have well defined analytical structures in an imaginary frequency domain when $\mu\neq \mu_c$ because all fermion fields are well gapped except at $\mu=\mu_c$ or TQCPs. 
If they are also invertible when $\mu \neq \mu_c$, they naturally form a representation of general linearized group ${\mathcal GL}(M, C)$ in $(d+2)$-dimension excluding  a singular point of $\mu=\mu_c$ and $\Omega={\bf k}=0$. 

Therefore, at a $(d+1)$ dimension surface of such a $(d+2)$ dimension hyperspace that excludes enclosed TQCPs (see Fig.1), $G(i\Omega, {\bf k};\mu)$ is smooth and can be thought as a mapping from a $(d+1)$ dimension surface to
$M \times M$ invertible complex matrices. 
This allows a classification of quantum fields in terms of Green's functions using $(d+1)$-dimensional sphere embedded in $(d+2)$ dimension hyperspace. Quantum fields can
be topologically non-trivial if the Green's functions $G(i\Omega, {\bf k}; \mu)$ at a given $\mu \neq \mu_c$ belong to nontrivial elements of $(d+1)$th homotopy group, i.e. $\pi_{d+1} ({\mathcal GL}(M, C))$.
This classification of the Green's functions had been emphasized in quite a few previous studies of gapped topological states\cite{Volovik03,Volovik09,Qi11,Gurarie11,Wang12}. It can also be applied to near TQCPs.
And if $G(i\Omega, {\bf k}; \mu>\mu_c )$ are associated with a different group element of $\pi_{d+1}({\mathcal GL}(M, C))$ compared with $G(i\Omega, {\bf k}; \mu <\mu_c)$, continuum quantum fields are in different topological classes across $\mu_c$
and therefore can be applied to study transitions involved changes of discrete topologies. This is a very unique feature of topological quantum criticality.  That is, effective quantum fields for  dynamics at TQCPs {\em
are representations of a topological group so to reflect changes of topologies of underlying many-body states}.

Simplest applications can be easily found in $2d$ time-reversal-symmetry breaking superconductors\cite{Read00}.
In this case, one finds that quantum fields for a TQCP form a nontrivial representation of ${\mathcal \pi}_{3}({\mathcal GL}(2,\mathbb{C}))=\mathbb{Z}$. 
At a surface of a fixed chemical potential $\mu (\neq \mu_c)$,  the topological invariant can be defined as ($k_0=\Omega$)

\begin{eqnarray}
N_W &= & \frac{1}{24\pi^2} \int d{k_x} dk_{y} dk_0 \epsilon_{\alpha\beta\gamma} \nonumber \\
&Tr& G \frac{\partial}{\partial k_\alpha} G^{-1}  G \frac{\partial}{\partial k_\beta} G^{-1}  G \frac{\partial}{\partial k_\gamma} G^{-1} , \alpha,\beta=0,x,y.
\nonumber \\
\label{NW}
\end{eqnarray}
This characterization can be extended to other even-spatial dimensions.

$N_W$ above can also be equivalently expressed in terms of the zero frequency limit of the Green's function, $G(0,{\bf k};\mu)$ without frequency integration.
Just as in the limit of mean-field or one-particle theory, one can then completely isolate the frequency integration from the momentum summation. 
The topological invariants above can then be more conveniently expressed in terms of $G^{-1}_{0}({\bf k},\mu)=G^{-1}(0,{\bf k}; \mu)$.
It is important to notice that although for a given $\Omega$, $G(i\Omega, {\bf k}; \mu \neq \mu_c)$ are complex $M \times M$ matrices and are non-Hermitian,
$G_0({\bf k},\mu)$ are Hermitian following the following general relation 

\begin{eqnarray}
G^\dagger (i\Omega, {\bf k};\mu) =G (-i\Omega, {\bf k};\mu).
\end{eqnarray}

At a fixed chemical potential,
$G_0({\bf k};\mu\neq\mu_c)$ thus defines a smooth mapping function from a $d$-momentum space to an effective Hamiltonian manifold, $H_M$ which topologically differs from ${\mathcal GL} (M,\mathbb{C})$.
This is most obvious if one takes the non-interacting limit when $G^{-1}_0({\bf k}) =H_0({\bf k})$. In the interacting limit, $G^{-1}_0({\bf k}; \mu \neq \mu_c)$ further contains effects of zero-frequency self-energies which are usually unknown
but the effective manifold $H_M$ shall be still be a hermitian one, the same as $H_0({\bf k})$.
So the topological charges above can also be conveniently applied to represent a nontrivial mapping of the homotopy group  ${\mathcal \pi}_{d=2}(H_M)$. 
For $2d$ time reversal symmetry breaking superconductors, $H_M$ is simply a two sphere $S^2$; and Eq.(\ref{NW}) also represents topological charges of smooth mapping function $G_0^{-1} ({\bf k};\mu \neq \mu_c)$ and reflects   
${\mathcal \pi}_2(S^2) =\mathbb{Z}$.

When applying to topological states with additional global symmetries such as time reversal symmetries in {\em odd} spatial dimensions, one has to further tweak the Green's function winding number definition to properly reflect global symmetries.
In this case, it is more convenient to directly work with the zero frequency Green's function, $G^{-1}(0, {\bf k};\mu)^{-1} =G^{-1}_0({\bf k}; \mu)$ that takes into account interaction renormalization effects.
Now to apply to $d$-dimensional spatial space of bulk states instead of $(d+1)$-dimensional space-time , we can again extend the space to $d+1$ dimension by inclusion of an additional dimension of the general tuning parameter $\mu$ along which a TQCP occurs. $G_0({\bf k}; \mu)$ defined in $(d+1)$ dimension hyperspace, has well defined analytical structures when $\mu\neq \mu_c$ because all fermion fields  are well gapped except at $\mu=\mu_c$ or TQCPs; they are also invertible 
when $\mu \neq \mu_c$ which is the situation we intend to focus on.

At a $d$-dimension surface of such a $(d+1)$ dimension hyperspace that again excludes enclosed TQCPs (see Fig.1), $G^{-1}_0({\bf k}; \mu \neq \mu_c)$ is everywhere smooth and can be thought as a mapping from $d$-dimension surfaces to
a Hamiltonian manifold. This allows a classification of quantum fields using a $d$-dimensional sphere embedded in $(d+1)$-dimension hyperspace. 
In the presence of time reversal symmetry in superconductors and therefore a chiral symmetry, $G^{-1}_0({\bf k}; \mu)$ can always be cast into an off-diagonal block matrix, 
with block matrices being $M \times M$ invertible complex matrices, $Q({\bf k})$ and $Q^\dagger({\bf k})$.  That is

\begin{eqnarray}
G^{-1}_0({\bf k}; \mu) &=& 
\begin{bmatrix}
0 & Q({\bf k})  \\
Q^\dagger ({\bf k}) & 0\\
\end{bmatrix}.
\label{Q}
\end{eqnarray}
This structure naturally appears in non-interacting or mean field theories and can be shown to be true in the presence of interactions as far as the chiral symmetry is present\cite{Volovik03,Volovik09,Qi11,Gurarie11,Wang12}.

$G^{-1}_0({\bf k};\mu \neq \mu_c)$ in Eq.(\ref{Q}) can be topologically non-trivial if their off diagonal complex matrices $Q({\bf k}; \mu)$ belong to nontrivial elements of {\em $d$-th homotopy group} of mapping functions, $\pi_d({\mathcal GL}(M, \mathbb{C}))$, instead of $\pi_{d+1}({\mathcal GL}(M, \mathbb{C}))$ of space-time Green functions discussed before. Note that $Q({\bf k};\mu)$ are $M\times M$ complex matrices while the space-time Green's function in this case, 
$G(i\Omega, {\bf k}; \mu)$, are $2M \times 2M$ invertible ones.

Simplest applications can be easily found in $3d$ time reversal symmetric $p$-wave superfluids or superconductors isomorphic to $^{3}He$ B-phase\cite{Leggett75}.
In this case, $M=2$ and quantum fields for a TQCP shall form
a non-trivial representation of ${\mathcal \pi}_{d=3} ({\mathcal GL}(2,\mathbb{C}))=\mathbb{Z}$; ${\mathcal GL}(2,\mathbb{C})$ now is the group for off-diagonal block complex matrices $Q({\bf k})$ appearing in Green's functions in the limit of zero frequency. The topological invariant can be defined in terms of off-diagonal block matrices $Q({\bf k};\mu)$ or formally in terms of zero frequency Green's function $G_0({\bf k};\mu\neq \mu_c)$ itself which is more suitable for
discussions of interacting fields.
In the later case, the winding number in the presence of time reversal symmetry (and hence chiral symmetry) defined in terms of $G_0({\bf k};\mu \neq \mu_c)$ needs to be supplemented by a chiral transformation $\Sigma$,

\begin{eqnarray}
N_W &=&\frac{1}{24\pi^2} \int d{k_x} dk_{y} dk_z \epsilon_{\alpha\beta\gamma}  \nonumber \\
&Tr& \Sigma G_0 \frac{\partial}{\partial k_\alpha} G_0^{-1}  G_0 \frac{\partial}{\partial k_\beta} G^{-1}_0 G_0 \frac{\partial}{\partial k_\gamma} G^{-1}_0, \alpha,\beta=x,y,z.
\nonumber \\
\label{NW1}
\end{eqnarray}
${\Sigma}={\mathcal T}{\mathcal C}$ is defined as a unitary chiral transformation anti-commuting with $G^{-1}_0({\bf k};\mu)$ that now is hermitian.
The winding number defined in Eq.(\ref{NW}) would vanish if $\Sigma$ were replaced with a unity matrix and hence it is crucial to have a chiral transformation $\Sigma$ in its definition.

\begin{eqnarray}
{ \Sigma}^{-1} G^{-1}_0({\bf k}) { \Sigma}&=&-G^{-1}_0({\bf k}). \nonumber \\
{\mathcal T}^{-1} G^{-1}_0({\bf k}) {\mathcal T}&=&G^{-1}_0(-{\bf k}),
{\mathcal C}^{-1} G^{-1}_0({\bf k}) {\mathcal C}=-G^{-1}_0(-{\bf k}). \nonumber \\
\end{eqnarray}
${\mathcal T}$, ${\mathcal C}$ are anti-unitary time reversal and charge conjugation transformation respectively.
In the case of $3d$ $p$-wave spin triplet superconductors, the effective Hamiltonian manifold $H_M$  suggested by hermitian matrix $G^{-1}_0({\bf k};\mu)$ can be defined on a three sphere $S^3$. So $N_W$  in Eq.(\ref{NW1}) also equivalently represents $\pi_{d=3}(S^3)=\mathbb{Z}$.

\begin{figure}
\includegraphics[width=8cm]{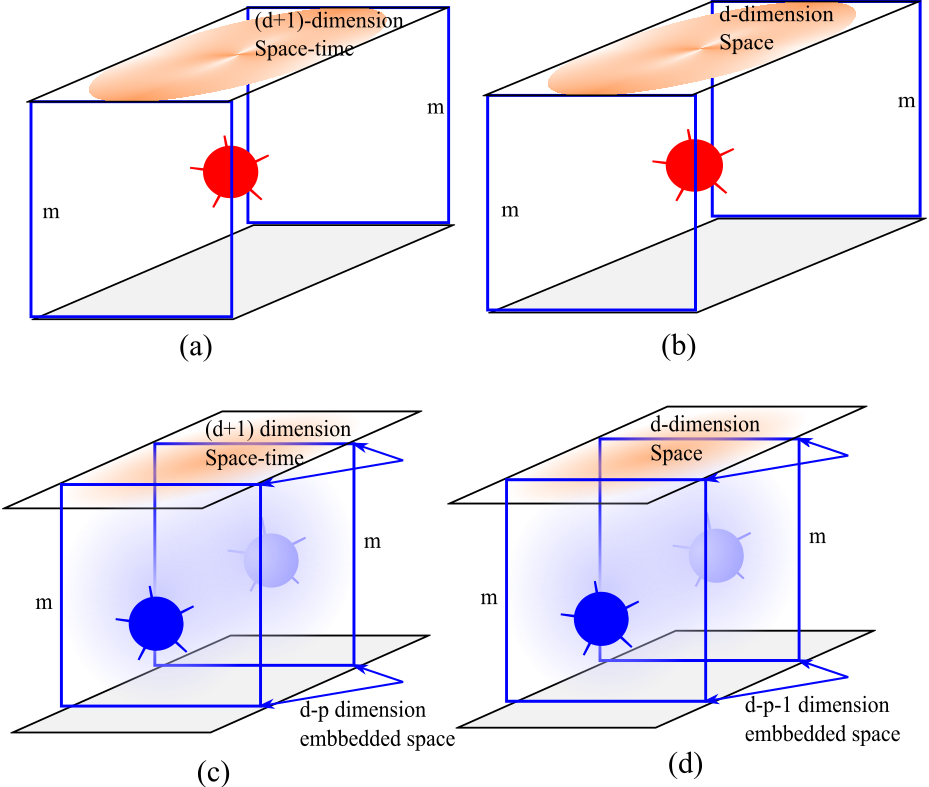}
\centering
\caption{ \footnotesize  Standard cartoon pictures of changes of global topologies (of integer groups); vertical direction represents a dimension of general tuning parameter $m$ for TQCPs. 
TQCPs (as red spheres) are between fully gapped topological superconductors (top) and trivial states (bottom) in a) and b). Majorana quantum fields form a representation of
either $\pi_{d+1}$homotopy group of space-time Green's functions in a $(d+1)$ dimension surface when $d$ is even (a)), or $\pi_d$ homotopy group of zero-frequency Green's functions in a $d$-dimension surface
embedded in $(d+1)$ dimension when $d$ is odd (b)). 
c) TQCPs between gapless topological superconductors and gapped states. For time reversal symmetry breaking TQCPs,
Quantum fields form a non-trivial class of $\pi_{d-p}$, the $(d-p)$th homotopy group of space-time Green's functions. $ d-p(>0)$ is the dimension of a subspace where topological invariants are defined.
$p=0,1$ are for a nodal point and nodal line phase respectively. In three dimension, only $p=0$ nodal points are stable.
d) is for time reversal invariant case. Quantum fields belong to non-trivial classes of  $\pi_{d-p-1}$, the $(d-p-1)$th homotopy group of zero-frequency Green's function or invertible complex matrix $Q({\bf k})$. 
Nodal points are stable only in $d=2$ but not in three spatial dimensions.}
\label{TQCP}
\end{figure}

For gapless topological phases, topological invariants are defined in an embedded subspace or sphere with dimension lower than the physical space-time or space dimensions.
A TQPC connects two states with different topologies in an embedded subsphere and again it crucially depends on the global time reversal symmetry.

For time reversal symmetry breaking states, the embedded space shall be a $[(d-p-1)+1]$ dimension space-time sphere transverse to a Fermi surface of dimension $p$, $p=0,1,...,d-1$..
Quantum fields for a TQCP shall be in a non-trivial representation of $[(d-p-1)+1]$th homotopy group of space-time Green's function for a fermi surface of dimension $p$.
For instance, for a time reversal symmetry breaking TQCP that results in emergence of a Fermi surface or nodal structure of $p$ dimension\cite{Yang21},
topological charges can be defined in a $(d-p)$-sphere or $S^{d-p}$. The space-time Green's functions shall be in a non-trivial class of homotopy group   ${\mathcal \pi}_{d-p} ({\mathcal GL}(M,\mathbb{C}))$ 
instead of ${\mathcal \pi}_{d+1}$ group for corresponding gapped topological states. And for nodal point phases, $p=0$ and one is concerned with ${\mathcal \pi}_{d}$ group of the space-time Green's function.
This homotopy group of superconducting Green's functions (i.e. $M \leq 2$) is indeed a non-trivial integer one, $\mathbb{Z}$ in $d=3$ indicating a stable time-reversal-symmetry breaking nodal point superconducting phase in three spatial dimensions.

In the presence of time reversal symmetry, again a subspace of $(d-p-1)$-spatial sphere transverse to a Fermi surface of dimension $p$ shall be introduced.
Zero frequency Green's function $G_0({\bf k}; \mu\neq \mu_c)$ can then be studied in a sphere of $(d-p-1)$-dimension and mapping functions shall be in 
a non-trivial class of $(d-p-1)$th homotopy group of the corresponding invertible smooth mapping matrices $Q({\bf k})$ in Eq.(\ref{Q}).
The Green's functions have to be in a non-trivial class of homotopy group of ${\mathcal \pi}_{d-p-1} ({\mathcal GL}(M,\mathbb{C}))$. For $d=2$, this group is a non-trivial integer group $\mathbb{Z}$ for nodal points ($p=0$) but
in $d=3$, only nodal lines ($p=1$) are topologically stable \cite{Beri10,Volovik03,Wen02,Kobayashi14}.

Quantum fields we are dealing with are continuous and do not have same ultraviolet structures as solid state Bloch bands, and don't have well-defined Chern numbers. However, by enforcing that they are in a topologically non-trivial sector of corresponding homotopy groups of  their Green's functions, one anticipates that continuum quantum fields are faithful representations of TQCPs for changes of discrete topology. 
A TQCP defined in this way then naturally connects two topologically distinct states in $d$-dimensions as illustrated in Fig.\ref{TQCP}.  
This was the key element behind previous constructions of effective field 
theories in quite a few different limits\cite{Yang19,Yang21}.  An important vindication for the validity of effective-field approaches to changes of global topologies in quantum states is predictions of topological boundary reconstructions across  a TQCP which one usually can verify quite straightforwardly by using its quantum field theories. This is what we now turn to in the next subsection.

\subsection{Bulk-Boundary Correspondence near TQCPs}

If two topologically different quantum fields 
are fused together in space, then spatial boundaries, or domain walls, separating them shall trap gapless (majorana if $U(1)$ symmetry is broken) boundary modes as a result of change of topologies in the physical space. 
Domain walls in a parameter space therefore directly lead to spatial domain walls or boundaries and such a phenomenon is robust and zero modes are usually stable as far as the underlying
protecting symmetries are respected and interactions are weak. This feature, a bulk-boundary correspondence, is also an essential one in discussions of gapped condensed matter topological states, either with $U(1)$ symmetry or with $U(1)$ symmetry broken.

As pointed out before, TQCPs are manifestation of changes of topologies. As we had been employing quantum field theories to facilitate many discussions of non-interacting  gapped topological states,
it is indeed not surprising and also natural to employ continunmm fields to explore interacting physics at gapless TQCPs if long wave length physics is the only aspect concerning us.
But there are at least three important caveats.

The first issue is related to definitions of topological invariants of topological states that, often, are only meaningful when protecting symmetries such as time-reversal symmetry are not broken.
As topologies might not be well defined when symmetries are broken by external fields, it has been confusing how gapped states evolve when symmetries are broken and whether resultant states shall still be (or not) related to a topological state. Continuum quantum fields offer a continuous view about what happens with or without protecting symmetries in the presence of interactions and provide very useful insight on resultant states.

To elaborate on this, we take an example that had been examined before. If a spin exchange or Zeeman field is introduced in a 3D time-reversal symmetric topological superconductor, the concept of topological invariants
defined explicitly in terms of Kramer doublets is no longer applicable. However, the effective field approach shows that i) there are {\em no bulk} transitions occur in small fields; ii) however, topological surfaces are critical at zero field
and are gapped in various ways when time reversal symmetries are broken explicitly; iii) bulk transitions occur at a finite Zeeman field exactly at the moment when surface states undergo complete reconstruction.
iv) the upper critical dimension for such TQCPs is $1\frac{1}{2}$, below 2d and interactions are irrelevant in 3d, implying a robust bulk-edge correspondence across TQCPs\cite{Yang21}. 
Aspect iii) is a principle feature of generic TQCPs. All bulk phase transitions at TQCPs occur exactly when topological surface states emerge or disappear, a bulk-boundary correspondence
that appears at all generic TQCPs we have examined so far. We have used it as another vindication of continumm field approaches to TQCPs.

The second caveat is related to protecting global symmetries at TQCPs and their roles in quantum dynamics of emergent quantum fields. We will discuss them in details in the nest section. 
The conclusion is that global symmetries set (minimal) numbers of fundamental fields that shall be present at gapless TQCPs. {\em  Numbers of emergent gapless fields at TQCPs} are {\em independent} of $N$, the number of solid state Bloch bands that are actually present in microscopic theories.  For TQCPs breaking all global symmetries including time-reversal ones and $U(1)$ symmetries, 
TQCPs can be effectively characterized by two-component  interacting majorana fields. For TQCPs with only time reversal symmetry
but breaking all other symmetries including global $U(1)$ symmetries as in topological superconductors, TQCP dynamics can be represented with four-component interacting majorana fields.
Although quantum phase transitions are not driven by local ordering and states on both sides of critical points can both break the same symmetry spontaneously, dynamics at TQCPs do crucially depend on global symmetries that remain unbroken at TQCPs.

Last issue is related to the second one but perhaps is more on strong coupling limits which we will elaborate in the next subsection.

\section{Strongly Interacting Limits}

Most discussions of solid state topological states, gapped insulators and superconductors have been inspired by non-interacting physics. 
So are our discussions on interacting TQCPs in the previous sections as well as discussions on the bulk-boundary correspondence above.
These discussions are expected to be valid at least when interactions are not very strong.

Whether all or some of above discussions on topological states can be extended from weakly interacting limits to certain strongly interacting limits
and to what extend are interesting questions and in general largely remain to be answered.
Most of this article is to deal this issue in a few broad classes of topological superconductors. Particularly, we exclusively focus on {\em continuous} evolutions of TQCPs when interactions of majorana fields allowed by global symmetries become strong. 

A very closely related question is what happens to bulk states when interactions increase.
One popular view is that as interactions increase in gapped phases, gapless boundary states can open up a gap without breaking protecting symmetries.
So one approach that had been taken is then to characterize interacting topological states by classifying topological ordered boundaries and apply the bulk-boundary correspondence to learn about bulk states. 
Those classifications provide general information on what are possible in extreme limits; on this, we refer readers to those original papers\cite{Fidkowski11,Fidkowski13,Metlitski15,Wang15,Song17}.
What kind of concrete interactions and interacting Hamiltonians can lead to those strongly interacting topological states are largely unknown to us.

Meanwhile,
gapless bulk quantum fields at TQCPs can also be strongly interacting although in high dimensions interactions are typically perturbative and irrelevant in the long wavelength limit. 
The possibility to extend discussions on TQCPs beyond weakly interacting limit is quite tempting. How TQCPs in superconductors in strong coupling limits differ from weak coupling limits is the key question we intend to answer in this article.

These discussions can further provide valuable detailed information on dynamics of gapped topological states near TQCPs in strongly interacting limits. 
So we anticipate gapless TQCPs are also excellent starting points to further discussions on interaction dynamic effects of gapped bulk states and limitations of various weakly-interacting pictures of TQCPs. We will return to this when discussing some strong coupling fixed point descriptions of TQCPs in superconductors and explicitly relate them to various new scenarios beyond more standard TQCPs of weakly interacting majorana fermions.

\section{The Role of Global Symmetries}

Let us now turn to the main focus of the article, the role of global symmetries.
We consider a general quantum phase transition in a topological state with $N(=2n)$ (including spins) continuum bands. More precisely, $N$ is the number of complex fermion degrees relevant to our discussions. 
A TQCP finds a representations in $2N=4n$-component majorana fields, $\chi$, with its effective Hamiltonian constructed in the following way,

\begin{eqnarray}
H &=& \int d{\bf r} \chi^T ({\bf r}) {\mathcal H}_0 \chi ({\bf r}) +H_I, \chi^\dagger(\bf r) =\chi^T(\bf r); \nonumber \\
{\mathcal H}_0 &=&{\mathcal P} +{\mathcal M}, {\mathcal P}^\dagger={\mathcal P} ={\mathcal P}^T, {\mathcal M}^\dagger ={\mathcal M}=-{\mathcal M}^T, \nonumber \\ 
\chi^T&=& (\chi_1, \chi_2,...,\chi_{4n}), \{ \chi_i({\bf r}), \chi_j({\bf r}') \}=\delta_{i,j} \delta({\bf r} -{\bf r}') \nonumber \\
\end{eqnarray}
Here the Hamiltonian matrix ${\mathcal H}_0$ is a $2N\times 2N$ hermitian matrix and $\chi({\bf r})$ are $4n$ component real fermions.
${\mathcal P}$ is odd in momentum or gradient operators and is real and symmetric for a given momentum. The mass matrix ${\mathcal M}$$(-\nabla^2)$ is even in momentum operators and is anti-symmetric, purely imaginary.
For a $N=2n$-band state, there are $N(2N-1)$ antisymmetric mass matrices and $N(2N+1)$ symmetric hermitian matrices. There are total $4N^2$ Hermitian matrices for a general construction.
So ${\mathcal M}$ can be of a general form of

\begin{eqnarray}
{\mathcal M} =\sum_{k=1}^{N(2N-1)} A_k {\Gamma}_k, {\Gamma}^\dagger_k ={ \Gamma}_k=-{ \Gamma}^T_k
\end{eqnarray}
where $\Gamma_k$, $k=1,...,N(2N-1)$ are anti-symmetric Hermitian matrices.
$H_I$ represents interactions among majorana particles or between majorana fields and other bosonic dynamic degrees of freedom; we will further specify its detailed structure in later sections.

For the purpose of our discussions, it is convenient to further arrange the hermitian $2N\times 2N$ mass matrix ${\mathcal M}$ in a form of block diagonal matrices after proper diagonalization.
${\mathcal M}$$({\bf p})$ in the limit of zero momentum ${\bf p}=0$ can therefore be conveniently expressed as a direct sum of diagonal mass matrices.

\begin{align}
{\mathcal H}_0 &= \begin{pmatrix} \begin{bmatrix}
\mathbb{P}_{11} & \mathbb{P}_{12} \\
\mathbb{P}_{21} & \mathbb{P}_{22}
\end{bmatrix}
-
\begin{bmatrix}
\mathbb{M}_{11} & 0 \\
0 & \mathbb{M}_{22}
\end{bmatrix}
\end{pmatrix} 
\end{align}
while in general ${\mathcal P}$ does not have a diagonal form and can not be expressed as a direct sum.

If TQCPs are time reversal invariant and if the time reversal symmetry is the only global symmetry, 
Hermitian anti-symmetric mass matrices ${\mathcal M}$ near TQCPs shall have the following generic form of direct sum

\begin{eqnarray}
{\mathcal M}({\bf p})&=&
\begin{bmatrix}
\mathbb{M}_{11} & 0 \\
0 & \mathbb{M}_{22}
\end{bmatrix}, 
\mathbb{M}_{22} =  \mathbb{M}_2 \oplus \mathbb{M}_3 \dots  \oplus \mathbb{M}_{N/2} ,\nonumber \\
\mathbb{M}_{11} ({\bf p} \rightarrow 0) &=&
 \begin{bmatrix}
m_1 & 0&0& 0 \\
0 & m_1 &0&0\\
0 & 0&-m_1 &0 \\
0 & 0& 0 &-m_1\\
\end{bmatrix}, \nonumber \\
\mathbb{M}_{i} ({\bf p} \rightarrow 0)&=&
 \begin{bmatrix}
m_i & 0&0& 0 \\
0 & m_i &0&0\\
0 & 0&-m_i &0 \\
0 & 0& 0 &-m_i\\
\end{bmatrix}, i=2,...,{N/2}  \nonumber \\ \nonumber \\
&& \mbox{so ${\mathcal M}({\bf p})$ is a diagonal bloch matrix} \nonumber \\
{\mathcal M}({\bf p}) &=&
 \begin{bmatrix}
\mathbb{M}_{11} & 0 &\dots &0 \\
0 & \mathbb{M}_2 &\dots &0 \\
\vdots & \vdots & \ddots &\vdots \\
0 & 0& \dots & \mathbb{M}_{\frac{N}{2}}  \\
\end{bmatrix} \nonumber \\
\label{mass}
\end{eqnarray}
In Eq.(\ref{mass}), there are total $N=2n$ pairs of eigenvalues of $\pm m_i$, $i=1,2,...,n=N/2$.
Without losing generality, we have ordered the eigen values as $0< m_1 < m_2 <...<m_{n}$ in the limit of ${\bf p}=0$.
Two fold-degeneracy in ${\mathcal M}$ reflects the time-reversal symmetry and the reflection symmetry in $\pm m_i$ follows the charge-conjugation symmetry of TQCPs considered here.

 If the time reversal symmetry is absent and no other global symmetries are present, the generic structure for the same $N$-band state shall be
 
 \begin{eqnarray}
{\mathcal M}({\bf p} )&=&
\begin{bmatrix}
\mathbb{M}_{11} & 0 \\
0 & \mathbb{M}_{22}
\end{bmatrix}, 
\mathbb{M}_{22} =  \mathbb{M}_2 \oplus \mathbb{M}_3 \dots  \oplus \mathbb{M}_{N}, \nonumber \\
\mathbb{M}_{11} ({\bf p} \rightarrow 0 )&=&
 \begin{bmatrix}
m_1 & 0 \\
0 &-m_1\\
\end{bmatrix}, \nonumber \\
\mathbb{M}_{i} ({\bf p} \rightarrow 0) &=&
 \begin{bmatrix}
m_i & 0\\
0 &-m_i\\
\end{bmatrix}, i=2,...,{N}  \nonumber \\ \nonumber \\
& & \mbox{Again ${\mathcal M}({\bf p})$ is a diagonal block matrix} \nonumber \\
{\mathcal M} ({\bf p}) &=&
 \begin{bmatrix}
\mathbb{M}_{11} & 0 &\dots &0 \\
0 & \mathbb{M}_2 &\dots &0 \\
\vdots & \vdots & \ddots &\vdots \\
0 & 0& \dots & \mathbb{M}_{{N}}  \\
\end{bmatrix}
\label{mass1}
\end{eqnarray}
where now they are $N$ pairs of non-degenerate eigenvalues $\pm m_i$, $i=0,1,..N-1$ ordered from the smallest to the largest.

For a discussion on TQCPs, it is important to keep in mind that the proper limit of ${\mathcal M}$ we need to take is 

\begin{eqnarray}
m_1 \rightarrow 0; \frac{m_i}{m_1} \rightarrow \infty \mbox{	for all $i \neq 1$.}
\label{QCP}
\end{eqnarray}
Eq.(\ref{QCP}) illustrates that only {\em one mass gap}, i.e. the smallest mass gap, closes while all other mass gaps remain open as in a generic quantum critical point.

Therefore, $\mathbb{M}_{11}$ can be used to projected out a low energy subspace for all TQCPs breaking time reversal symmetry. More important, it is the dimension of $\mathbb{M}_{11}$
not that of the full mass matrix ${\mathcal M}$ that defines effective field theories for TQCPs.
In other words, $\mathbb{M}_{11}$ effectively defines a fundamental fermion representation of global symmetries. 
The dimension of ${\mathcal M}_{11}$ is $2\times 2$ for time-reversal symmetry breaking 
TQCPs.The simplest example in this class can  a 2D $p$-wave $Z_2$ topological superconductors with $N=2$ in the presence of a Zeeman field.  Breaking time reversal symmetry results in 
$\mathbb{M}_{11}$ above and effectively projects out  a time reversal symmetry breaking $p+ip$ theory.

For time reversal invariant TQCPs,
${\mathbb M}_{11}$ is minimally $4\times 4$ and the minimum can be reached only when there are no any additional global symmetries. 
The simplest example is a 3D time reversal odd-party p-wave topological superconductor that is in a minimal representation to start with. We anticipate other topological superconductors breaking all global symmetries except the time reversal symmetry can be described by the same effective field theory.

In both cases, the effective mass operators can be much smaller than $2N \times 2N$ mass matrices ${\mathcal M}$ of the original $N$-band complex fermions.  
This is the one of main results of this article. We have found that the dimension of effective fermion fields for TQCPs are largely independent of $2N$, the number of microscopic real fermion bands involved, but only depends on
global symmetries present at TQCPs, a quite remarkable feature of topological quantum criticality. Below we will show that strongly coupling limit physics crucially depends on global symmetries because of the mass operators discussed here.

All discussions on time reversal symmetric TQCPs can be easily generalized to TQCPs with other global symmetries as far as mass operators can be reduced to the same structure shown above.
In this sense, time reversal symmetry is not essential for our discussions although in this article we have chosen to work with a situation where the only single global symmetry present at TQCPs happens to be the time reversal one.

To illustrate this point further, we introduce two projection operators using a direct sum form,

\begin{eqnarray}
&& P_+ ={\mathbb{I}_{l\times l}} \oplus \mathbb{O}_{(2N-l)\times (2N-l)}, \nonumber \\ 
&&P_+ \cdot P_-=\mathbb{O}_{2N\times 2N}, P_+ +P_-=\mathbb{I}_{2N\times 2N}. \nonumber \\
\end{eqnarray}
where $l=4$ for time reversal symmetric TQCPs and $l=2$ for time reversal symmetry breaking TQCPs. 
$\mathbb{I}$ and $\mathbb{O}$ are unity and null matrices respectively.
This is to be consistent with the dimension of $\mathbb{M}_{11}$ that has been defined in Eq.(\ref{mass}),(\ref{mass1}) with and without global symmetries respectively.

Effective majorana fields, two- or four-component ones projected out by these operators $P_{\pm}\chi({\bf r}) =\chi_{\pm}$ shall have the following dynamics in the limit of defined in Eq.(\ref{QCP}),

\begin{eqnarray}
H_{eff} &=& =\int d{\bf r} \chi^T_+ ({\bf r}) {\mathcal H}_{e0} \chi_+ ({\bf r}) +H_{eI}, \chi_+^\dagger(\bf r) =\chi_+^T(\bf r); \nonumber \\
{\mathcal H}_{e0} &=&\mathbb{P}_{11}({\bf p})  -\mathbb{P}_{12} ({\bf p}) \mathbb{M}^{-1}_{22} ({\bf p}) \mathbb{P}_{21}({\bf p})+\mathbb{M}_{11}({\bf p})+ \dots\nonumber \\
\label{Heff}
\end{eqnarray}
where ${\bf p}=-i\nabla$ and the second term in ${\mathcal H}_{e0}$ that is even in momenta shall be less relevant in the long wavelength limit than the first one which can be linear in momenta.

The Lorentz symmetry can naturally emerge for real fermions in the low energy limit near TQCPs which we will further explore below. 
If an emergent Lorentz symmetry appears at a TQCP, ${\mathcal P}_{11}$ and ${\mathbb{M}}_{11}$ shall have the following structures,

\begin{eqnarray}
&& {\mathbb{P}}_{11} = -i\nabla_\alpha\gamma_\alpha, \mathbb{M}_{11}=\gamma_0,  \gamma_\alpha^\dagger =\gamma_\alpha=\gamma_\alpha^T, \gamma_0^\dagger=\gamma_0 =-\gamma^T_0 \nonumber \\
&&\{ \gamma_\alpha, \gamma_\beta \}=\delta_{\alpha\beta}, \{\gamma_0,\gamma_0\}=1, \{\gamma_\alpha,\gamma_0 \}=0; \alpha=1,..,d  
\label{lorentz}
\end{eqnarray}
where $\gamma_{\alpha,0}$, $\alpha=1,...,d$ are anti-communting hermitian matrices.
Furthermore, $\gamma_\alpha$, $\alpha=1,..,d$ are real and symmetric while $\gamma_0$ is purely imaginary and anti-symmetric.
Obviously, for time reversal symmetry breaking cases with 2-component majorana fermions projected out by mass matrices ${\mathbb{M}}_{11}$,
$\gamma_{i,0}$ shall be $2\times 2$ ones and can only be constructed in two spatial dimensions or $d=2$ but not in $d=3$.
With time reversal symmetry, $\gamma_{i,0}$ shall be $4\times 4$ ones and can be constructed in both two and three spatial dimensions or $d=2,3$.

In general, Eq.(\ref{lorentz}) can not be satisfied and there will be no emergent Lorentz symmetry. Instead, generic mass terms can further break space-time Lorentz symmetry or even spatial rotational symmetries leading to
quantum Lifshitz majorana fields (QLMFs). Eq.(\ref{Heff}) can also be applied to TQCPs characterized by QLMFs; these TQCPs include transitions into various topological nodal phases. 
A few concrete limits had been examined before in Ref.\cite{Yang21} and Eq.(\ref{Heff}) is the most general description of those dynamics.

Before concluding this section, we want to further remark on possible extensions to other global symmetries that can be present in solid states.
If more global symmetries are present, we expect that the size of relevant mass matrix ${\mathbb{M}_{11}}$, based on which $P_{+}$ is defined, becomes larger while the size of ${\mathcal M}$ is fixed.
More symmetries would lead to more emergent fermion fields at TQCPs and more components of majorana fields. 
This was also observed in a few concrete models studied in Ref.\cite{Yang21} and we do not pursue further in this presentation.

Below we will discuss some manifestation of global symmetries in energetics and dynamics.




\section{TQCPs near Strong Coupling Fixed Points I}

Effective quantum field theories suggest that upper critical dimensions for most TQCPs are usually lower than 2d so in high dimensions that our studies are focused on, generic TQCPs are represented by 
weakly interacting majorana fields.
This leads to free majorana fermion universality classes and results in higher order quantum phase transitions, typically higher than the second-order Landau phase transitions.
For TQCPs with Lorentz symmetry, transitions are $(d+1)$th order ones for $d=2,3$\cite{Yang19}.
Weak interactions also suggest stability of topological surface states and robustness of surface-bulk correspondence away from these TQCPs.

However, if majorana fields in a particular physical system happen to be strongly interacting, then TQCPs in this limit can be strong coupling and there can be higher emergent symmetries,
higher than the simple generic scale or scale-conformal symmetries that usually appear at critical points. We are now turning to such limits. We will demonstrate below that these higher emergent 
symmetries, or more specifically various supersymmetries, also crucially depend on global symmetries discussed above. 

When there are no global symmetries present at a TQCP and time reversal symmetries
are broken, at strong coupling fixed points majorana fermions alway posscess an emergent supersymmetry with one supercharge, i.e. single (two-component) majorana fermion with one emergent real scale field; they belong to the simples 
${\mathcal N}=4N_f={1}$  supersymmetry theory. $N_f$ refers to the number of fermion degrees of freedom in term of $(3+1)D$ Dirac fermions and ${\mathcal N}$ is the number of two-component real fermions. In other words, there is only one single strong coupling fixed point in this limit.
Supersymmetry can also naturally emerge in strongly correlated nodal phases\cite{Lee07,Balents98}, on topological surfaces\cite{Grover14,Ponte14,Li17} or in other interesting correlated systems\cite{Yu10,Jian17,Zerf16,Fendley05,Rahmani15,Affleck17}. 
Here we have found that the supersymmetry naturally emerges at strong coupling fixed-points of TQCP dynamics if {\em time reversal symmetry is broken}. 

When a global symmetry is present and say a TQCP is time reversal symmetric, there are multiple strong coupling fixed points. Usually, they only involve one emergent real scalar field and are non-supersymmetric.
However, when two real scale fields are emergent in dynamics and both interact with majorana fermions, the strong coupling fixed points of majorana fermions can then be supersymmetric and belong to $N_f=\frac{1}{2}$ class.
These discussions are based on properties of interacting gapless majorana fermion fields and so are also valid in the context of gapless topological states. Global symmetries can be applied to classify strongly interacting gapless phases.

\subsection{Time Reversal Symmetry Breaking TQCPs in $2d$ I: Local interactions}

Let us start with the simplest limit of TQCPs when time reversal symmetry as well as all other symmetries are all broken.
The effective theory in $2d$ with a local and rotational invariant interaction can be expressed in the following representation;

\begin{eqnarray}
H &=& \int d{\bf r} [  \chi^T ({\bf r}) ( \tau_x  i\nabla_z  -\tau_z  i\nabla_x )
\chi ({\bf r}) \nonumber \\
&+& g\chi^T  \tau_y \frac{\nabla}{i} \chi ({\bf r}) \cdot \chi^T \tau_y \frac{\nabla}{i} \chi({\bf r}) ]; \nonumber \\
\chi^T&=& (\chi_1, \chi_2), \{ \chi_i({\bf r}), \chi_j({\bf r}') \}=\delta_{i,j} \delta({\bf r} -{\bf r}'), i=1,2. \nonumber \\
\label{H1}
\end{eqnarray}
where $\nabla=(\nabla_x,\nabla_y)$. Note that because only two-component majorana fermions emerge in this limit, terms of a form of $\chi^4$ without involving gradient operators vanish identically because of fermion statistics.
The four-fermi operator in Eq.(\ref{H1}) is the most relevant one for your discussions.

The Hamiltonian for TQCPs in Eq.(\ref{H1}) displays an emergent $Z_2$ symmetry when $\chi$ transforms as

\begin{eqnarray}
\chi_{1} \rightarrow K \chi_{2}, \chi_{2} \rightarrow - K \chi_{1},
\label{Z21}
\end{eqnarray}
where $K$ is an action of complex conjugate. 
Under such a transformation, $\tau$ matrices in Eq.(\ref{Z21}) transform accordingly $\tau_{x,z} \rightarrow  - \tau_{x,z}$, $\tau_y\rightarrow  -\tau_y$.
The interacting Hamiltonian for TQCPs is invariant under the transformation in Eq.(\ref{Z21}). {\em This symmetry only emerges at TQCPs and is not a symmetry of adjacent gapped topological states as a mass term breaks such a symmetry.}

The emergent $Z_2$ symmetry differs from the time reversal symmetry which is broken in our case.
Time reversal transformation is defined in terms of  
\begin{eqnarray}
\chi_{1} \rightarrow K \chi_{1}, \chi_{2} \rightarrow - K \chi_{2}.
\label{TRB2}
\end{eqnarray}
Accordingly, $\tau_x \rightarrow -\tau_x$ and $\tau_{z,y} \rightarrow \tau_{z,y}$ and the TQCP Hamiltonian in Eq.(\ref{H1}) is not invariant under this transformation as stated before.

Under scale transformation, $g$ follows the following renormalization group equation in  terms of dimensionless coupling  constant $\tilde{g}=c_d g \Lambda^{d+1}$,where $c_d$ is a constant of order of unity  and in $d=2$, $c_2=\frac{1}{4\pi^2}$.

\begin{eqnarray}
\frac{d\tilde{g}}{dt} =(d+1)\tilde{g} + \tilde{g}^2, t=\ln \Lambda;
\label{RGE0}
\end{eqnarray}
and we will mainly apply to $d=2$ in our discussions as time reversal symmetry breaking topological superconductors only exist in $2d$.
Here $\Lambda$ is a running ultraviolet scale in the scale transformation ad $c_d$ is a constant of order of unity and Eq.(\ref{RGE0}) is obtained in a one-loop approximation.

As suggested in Eq.(\ref{RGE0}), the free majorana fermion fixed point with $\tilde{g}=0$ is infrared stable in $d=2$ as the scaling dimension of interaction operator is $3$ (and scale dimension $0$ is for a marginal interaction).
It shows that as far as $\tilde{g}$ is small, weakly interacting gapless majorana fermions can form a well defined phase. So $(d+1)$th order transitions for $d=2$ time reversal symmetry breaking TQCPs \cite{Yang19} that belong to free majorana fermion universality class shall be very robust  as weak interactions are highly irrelevant. 

Although technically speaking Eq.(\ref{RGE0}) is perturbative, it is still suggestive to extrapolate to strong coupling limits and to gain some insight about what happens when $\tilde{g}$ is not too small.
Eq.(\ref{RGE0}) dose imply a strong coupling fixed point with $\tilde{g}^*=-(d+1)$. 
It appears that as far as $\tilde{g} > \tilde{g}^*$,  the free fermion fixed point at $\tilde{g}=0$ remains infrared stable, leading to a well-defined weakly interacting gapless majorana phase boundary. 
The corresponding transitions of this universality class is of $(d+1)$th order in $d=2$ as pointed out in a previous article\cite{Yang19}.  

However, in the strong coupling limit where $\tilde{g} < \tilde{g}^*=-(d+1)$, $\tilde{g}$ can follows into $\tilde{g}^*=-\infty$.
This indicates a distinct strong coupling phase of majorana fermions where a mass gap of either sign, $m$ or $-m$ is spontaneously created and fermions are fully gapped.
The ground state therefore shall be two-fold degenerate corresponding to two masses $m$ and $-m$ and spontaneously break the emergent $Z_2$ symmetry at TQCPs.

The strong coupling phase suggested above is an analogue of Gross-Neveau mass generation well understood in quantum field theories\cite{Gross74,Fei16}.
Indeed, if one further introduces $N$ flavours of majorana fermions that interact in an identical way as shown in Eq.(\ref{H1}), then in the limit of large $N$, one can obtain a renormalization equation similar to Eq.(\ref{RGE0}) above.
One can show that in the leading order of $(1/N)$, Eq.(\ref{RGE0}) can be applied to describe renormalization of a coupling constant $\tilde{g} =c_d N^2 g$ even when $\tilde{g}$ is much larger than unity. So in the large $N$ limit, the strong coupling physics conjectured above can be easily verified.

The strong coupling phase suggested above can be further connected to strongly interacting lattice models of majorana fermions carefully studied before\cite{Rahmani15, Affleck17}.
In $1d$, it had been shown explicitly that strongly interacting fixed points implied in Eq.(\ref{RGE0}) with $\tilde{g}^*=-(d+1)$ is directly related to a supersymmetric tri-critical Ising point observed and confirmed in the lattice model. 

Therefore, implications of the simple one-loop renormalization equation, Eq.(\ref{RGE0}) are highly illuminating, shedding light on TQCPs in strong coupling limits. It suggests that the weakly interacting $(d+1)$th order transition line be terminated by
a strong coupling fixed point beyond which there shall be discontinuous first order transitions.
To better understand strong coupling physics of TQCPs without time reversal symmetry, and strong coupling fixed points, below we introduce an extended model that is more suitable for current studies. 
The model below can be viewed as an ultraviolet completion of the infrared physics discussed so far. It not only leads to the same infrared physics discussed above but also naturally extends the theory to further higher inmtermediate energy scales which is fully consistent with the infrared physics.

\subsection{Time Reversal Symmetry Breaking TQCPs in $2d$ II: General discussions}

Near a strong coupling fixed point and beyond, the long wave length physics is expected to be naturally and more explicitly described by majorana fermions interacting with emergent scalar fields.
Therefore, to pursue discussions in strong coupling limits, it is more convenient to introduce the following extended model that is more suitable for such a purpose.

\begin{eqnarray}
H &=& H_{0m} +H_{0s}+H_I, \nonumber \\
H_{0m} &=&  \int d{\bf r} [  \chi^T ({\bf r}) ( \tau_x  i\nabla_z  -\tau_z  i\nabla_x )
\chi ({\bf r})
\nonumber \\
H_{0s} &=&\int d{\bf r} [ \pi^2({\bf r})  + \nabla \phi ({\bf r}) \cdot \nabla \phi({\bf r})  + M^2 \phi^2({\bf r}) ]; \nonumber \\
H_I &=& \int d{\bf r} [g_Y \phi({\bf r}) \chi^T({\bf r}) \tau_y \chi({\bf r}) +g_4 \phi^4({\bf r})];  \nonumber \\ \nonumber \\
& [ \phi({\bf r}) &,  \pi({\bf r}') ]  = i\delta({\bf r}-{\bf r'}), [\phi({\bf r}),\phi({\bf r'}]=[\pi({\bf r}),\pi({\bf r}')]=0. \nonumber \\
\label{SUSY1}
\end{eqnarray}
where

\begin{eqnarray}
\chi^T= (\chi_1, \chi_2), \{ \chi_i({\bf r}), \chi_j({\bf r}') \} =\delta_{i,j} \delta({\bf r} -{\bf r}'), i=1,2. \nonumber \\
\end{eqnarray}

The Hamiltonian, though breaking the time reversal symmetry, can have a $Z_2$ symmetry; i.e. Eq.(\ref{SUSY1}) is invariant under the following transformation

\begin{eqnarray}
\chi_{1} \rightarrow K \chi_{2}, \chi_{2} \rightarrow - K \chi_{1},
 \phi \rightarrow  -\phi, \pi \rightarrow -\pi
\label{Z2}
\end{eqnarray}
 where $K$ is an action of complex conjugate. Again, under such a transformation, $\tau_{x,z} \rightarrow - \tau_{x,z}$ and $\tau_y \rightarrow - \tau_y$.
 The transformation is effectively a $\pi$ rotation around the $\tau_y$ axis followed by an operation of $K$.

The $Z_2$ symmetry defined above differs from the time reversal one as in the later case, $\phi, \chi$ transform differently as

\begin{eqnarray}
\chi_{1} \rightarrow K\chi_1, \chi_2 \rightarrow - K\chi_2, \phi \rightarrow \phi, \pi \rightarrow \pi.
\label{TRB}
\end{eqnarray}
As stated before, the Hamiltonian in Eq.(\ref{SUSY1}) breaks the time reversal symmetry and is not invariant under the transformation in Eq.(\ref{TRB}).

$Z_2$ symmetry in Eq.(\ref{Z2}) can be more easily understood if one integrates out fermions first and constructs an effective potential for $\phi$ field.
The effective potential is manifestly invariant under a standard reflection of a real scalar field, $\phi \rightarrow -\phi$. The ground state is $Z_2$ invariant if $\langle \phi \rangle=0$ but otherwise is two-fold degenerate if spontaneous symmetry 
breaking occurs and $\langle \phi \rangle \neq 0$. {\em Note that this is an emergent symmetry of interacting gapless majorana fermions exactly at TQCPs but is not a symmetry of adjacent gapped topological states}. One can verify that a gapped topological state which is non-degenerate breaks this $Z_2$ symmetry explicitly.

When $M^2 $ is positive and finite, the ground state is non-degenerate and is $Z_2$ symmetric as $\langle \phi \rangle=0$; majorana fermions remain massless. 
One can further integrate over the gapped scalar fields $\phi$ and generate four fermion interactions in $\chi$ fields similar to interactions in Eq.(\ref{H1}). In the infrared limit (when momentum scale $\Lambda \ll M$) , fermion interactions mediated by the gapped scalar fields do have a simple  form of

\begin{eqnarray}
H_I= \frac{g^2_Y}{2M^4} \int d{\bf r} \nabla ( \chi^T({\bf r}) \tau_y \chi({\bf r})) \cdot \nabla (\chi^T({\bf r}) \tau_y \chi({\bf r}))+...
\nonumber \\
\label{Induced}
\end{eqnarray}
where we have muted higher order terms or other terms not relevant to the following discussions. Again because of fermion statistics, terms like $\chi^4$ without involving $\nabla \chi$ vanish identically and so there are no contributions of order of $O(1/M^2)$. 
Eq.(\ref{Induced}) can be easily related to $H_I$ in Eq.(\ref{H1}) if one sets  

\begin{eqnarray}
g \sim -\frac{2g^2_Y}{M^4}.
\end{eqnarray}

Therefore, majorana fermions interacting with massive scalar fields $\phi$ as shown in Eq.(\ref{SUSY1}) are naturally related to interacting gapless majorana fermions represented in Eq.(\ref{H1}).
The limit with a large mass gap $M^2 (>0)$ corresponds to the weakly interacting limit with small $g$. And at any finite mass gap $M^2$, the long wavelength limit of Eq.(\ref{SUSY1}) is hence expected to be equivalent to Eq.(\ref{H1}). 
In the infrared limit, we anticipate that the model with a finite mass $M^2$ again follows into the infrared stable free majorana fixed point with gapped scale fields that play little role in low energies.
So Eq.(\ref{SUSY1}) is an extended model of Eq.(\ref{H1})  and is more convenient for discussions of strong coupling TQCPs.

\begin{figure}
\includegraphics[width=6cm]{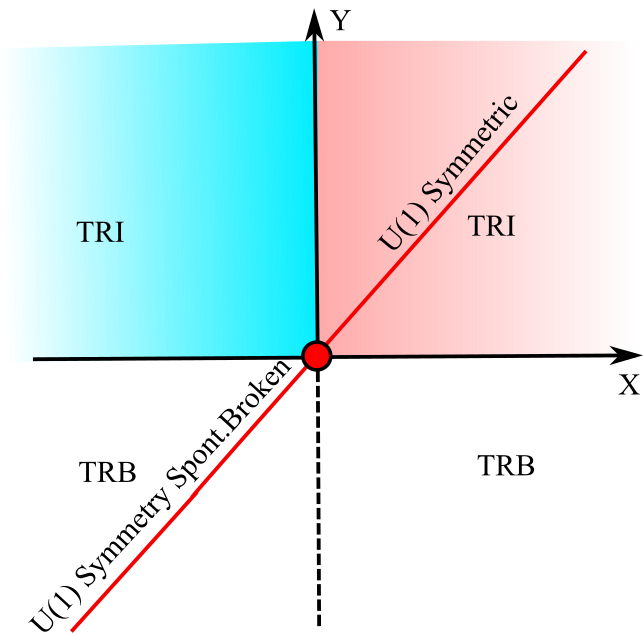}
\centering
\caption{ \footnotesize  Phase diagrams of $Z_2\otimes Z_2$ (for ${\mathcal T}$ and ${\mathcal R}$) symmetric model in Eq.(\ref{SUSY3}) in a plane of $X=M_1^2$, $Y=M_2^2$.
Upper right corner in this digram is weakly interacting.
In  the half plane of $Y>0$, states are time reversal invariant (TRI) while in the half plane of $Y<0$, time reversal symmetry is broken spontaneously (TRB). In the quadrant of
$X < 0, Y>0$,  $Z_2$-reflection symmetry is broken while in $X>0, Y>0$, states are $Z_2 \otimes Z_2$ symmetric.
In $X>0, Y<0$, $Z_2$ symmetries of both ${\mathcal T}$ and ${\mathcal R}$ are broken but the combined symmetry of ${\mathcal TR}$ remains;
 $Z_2 \otimes Z_2$ is spontaneously broken down to $Z_2$ for combined ${\mathcal T}$ and ${\mathcal R}$ transformation. 
$X=0, Y > 0$ and $Y=0, X > 0$ lines are for $Z_2 \otimes Z_2$ symmetric {\em CFT} states of {\em Gross-Neveu} fixed points or {\em GNFP} classes in $2d$. 
In this plane, only $X=Y > 0$ line is $U(1)$ symmetric if $g_{Y1}=g_{Y2}$ (see Eq. (\ref{SUSY3})). $U(1)$ symmetry is broken spontaneously along $X=Y <0$ and broken explicitly in the rest of plane. Point $X=Y=0$ in $2d$ is supersymmetric with $N_f=\frac{1}{2}$ and $U(1)$ symmetry;
in $d=3$, it is infrared trivial.
For $Y>0$, TQCPs are either $(d+1)$th order ($X>0$) or first order ones ($X<0$). For $Y<0$, there shall be no transitions because of spontaneously time reversal symmetry breaking in this limit of strong interactions.}
\label{U(1)}
\end{figure}

An interesting limit of Eq.(\ref{SUSY1}) is when  $M^2$, the mass gap of scalar fields also vanishes resulting in a scale invariant strong coupling state. 
When $M^2=0$ in dimensions lower than $3d$ including $2d$, non-interacting fixed point with $\tilde{g}^*_Y=0$ becomes infrared unstable.
The only infrared stable fixed point in $2d$ is strong coupling and is again supersymmetric\cite{Fei16,Grover14,Zerf16}.

The scale symmetric solution to Eq.(\ref{SUSY1}) when $M^2=0$ shall be identified as an effective description of the anticipated strong coupling fixed point $\tilde{g}^*$ discussed in the previous subsection. This fixed point represents a conformal-field-theory ({\em CFT}) state with an additional supersymmetry. Furthermore,
the strong coupling fixed point is also $Z_2$ symmetric.
 
When $M^2$ further becomes negative, there is standard spontaneous symmetry breaking of $Z_2$ symmetry in real scalar field $\phi$ and $\langle \phi \rangle \neq 0$. This leads to a mass generation for majorana fermions, $m \sim g_Y \langle \phi \rangle$ and two-fold degenerate ground states in the strong coupling limit. We identify this as the strongly coupling limit of gapless majorana fermions.
By varying the mass gap $M^2$, we anticipate that Eq.(\ref{SUSY1}) can be applied to capture both weak coupling and strong coupling limits suggested by the renormalization group equations of Eq.(\ref{H1}). And
Eq.(\ref{SUSY1}) with $M^2=0$ shall be an effective representation of the strong coupling point $\tilde{g}^*$ that turns out to be conformal with an emergent supersymmetry.

We can apply these results of interacting gapless majorana fermions to examine strong coupling TQCPs.
In $2d$, the strong coupling fixed point above indicates that the $(d+1)$th order transition line obtained in the weakly interacting limit shall be terminated at $\tilde{g}^*$ with an emergent supersymmetry. Beyond this point, a majorana mass, either positive or negative is spontaneously generated by interactions. The mass $m$ is a function $(\tilde{g}-\tilde{g}^*)$ and ground state is two-fold degenerate. We identified this as a first order transition line (See Fig. \ref{SUSY}). 

By passing, we want to point out that the field theory model itself can be easily reduced to $d=1$ dimension which can be applied to study {\em spinless} time reversal symmetric TQCPs in $1d$.
The free fermion fixed points are again stale. So the continuous $(d+1)$th or $2$nd order transition lines in this case is again terminated by a supersymmetric fixed point beyond which first order transitions occur.
The conclusions here are qualitatively identical to $2d$ time reversal symmetry breaking TQCPs.

\section{ TQCPs near Strong Coupling Fixed Points II} 

\subsection{Time reversal invariant case I: Local interactions and emergent $U(1)$ symmetry}

With a single global symmetry, say time reversal symmetry, at a TQCP, the minimal representation becomes a four-component majorana fermion field. If there are no other global symmetries at TQCPs, only minimal four majorana fields shall emerge at low energies relevant to our discussions.
 A generic local interacting Hamiltonian in $3d$ with Lorentz symmetry can have the following form,

\begin{eqnarray}
H &=& \int d{\bf r} [ \chi^T({\bf r}) (\tau_z \otimes (\sigma_x {i\nabla_z} -\sigma_z {i\nabla_x}) +\tau_x \otimes \mathbb{I} {i\nabla_y}) \chi ({\bf r}) 
\nonumber \\
&+& g_1 \chi^T({\bf r}) \tau_y \chi({\bf r}) \chi^T({\bf r}) \tau_y \chi({\bf r}) ]; \mbox{}
\chi^T= (\chi_{1\uparrow}, \chi_{1\downarrow}, \chi_{2\uparrow},\chi_{2\downarrow}). \nonumber \\
\label{TQCPTRI}
\end{eqnarray}
Indeed, Eq.(\ref{TQCPTRI}) can also be an effective  theory for a topological superconductor with p-wave pairing. 
One can reduce Eq.(\ref{TQCPTRI}) to $2d$ or $1d$ by muting one or two spatial gradient operators.

It is important to remark that if there are no other less relevant local interactions involving gradient operators,
Eq.(\ref{TQCPTRI}) as a model for TQCPs has a hidden $U(1)$ symmetry although the $U(1)$ symmetry associated with microscopic fermion conservation is spontaneously broken because of superconductivity.
{\em And this emergent symmetry only appears exactly at TQCPs but is not present in gapped phases adjacent to TQCPs.}

To further visualize the symmetry, it is useful to introduce a unitary rotation 

\begin{eqnarray}
& & U(\frac{2\pi}{3})=\frac{1}{2} [1+i (\tau_x \otimes \sigma_y +\tau_y+\tau_z \otimes \sigma_y)] \nonumber \\ 
& & U^\dagger (\tau_x \otimes \sigma_y, \tau_y, \tau_z \otimes \sigma_y) U
\rightarrow (\tau_y, \tau_z \otimes \sigma_y, \tau_x \otimes \sigma_y). \nonumber \\
\label{UT}
\end{eqnarray}
Correspondingly, $\chi \rightarrow U \chi$.
Now in terms of $g_0=g_1/2$, 
$H(\{\chi\}) \rightarrow H' (\{\chi\})$ and $H'$ has the following form
\begin{eqnarray}
H'&=& \int d{\bf r} [ \chi^T ({\bf r}) (\mathbb{I}  \otimes (\sigma_x  {i \nabla_x} +\sigma_z {i\nabla z}) +\tau_y \otimes \sigma_y  {i\nabla_y}) \chi ({\bf r}) 
\nonumber \\
&+& g_0 \chi^T \tau_x \otimes \sigma_y \chi \chi^T \tau_x\sigma_y \chi +g_0 \chi^T \tau_z \otimes \sigma_y \chi \chi^T \tau_z \otimes \sigma_y \chi ]; \nonumber \\
\label{SUSY2}
\end{eqnarray}
Eq.(\ref{SUSY2}) is explicitly invariant under any rotation around $\tau_y$ and therefore actually has an emergent $U(1)$ symmetry. 
It suggests that this shall be a unique property of interacting majorana fermions with local four-fermion operators (without involving gradient operators).

Note that for the convenience of later presentations and for the transparency, in deriving Eq.(\ref{SUSY2}), we have explicitly employed two different representations for the local four-fermion interaction operator in Eq.(\ref{TQCPTRI}). 
Since we deal with four component majorana fermions,
one can easily further show
that the two four-fermion operators in Eq.(\ref{SUSY2}) are actually identical and both are generated from a single four fermion operator $g_0 \chi^T \tau_y \chi \chi^T \tau_y \chi$.

We also have found that this tree level symmetry is respected when further quantum corrections are taken into account.
The one-loop renormalization group equation for $g_{0}$ in Eq.(\ref{TQCPTRI}) can be easily obtained via introducing $\tilde{g}_0(\Lambda)=c_d g_0 \Lambda^{d-1}$ where for $d=2$, $c_2=\frac{1}{4\pi^2}$;

\begin{eqnarray}
\frac{d\tilde{g}_0}{dt} =(d-1)\tilde{g}_0 + \tilde{g}_0^2, t=\ln \Lambda
\label{RGE1}
\end{eqnarray}
where $\Lambda$ is a running ultraviolet scale for the interacting Hamiltonian.
The free fermion fixed point $\tilde{g}=0$ above is stable in $d=2,3$ where main applications had been found\cite{Yang19};
$(d+1)$th order transition lines of this universality class therefore shall persist in the weakly interacting limit.

Eq.(\ref{RGE1}) itself also has strong coupling fixed points, $\tilde{g}^*_0=-(d-1)$ in $2d$ and $3d$. Below we will mainly focus on $2d$ which is most interesting.
When $\tilde{g} \geq \tilde{g}^*$, we do not expect spontaneously breaking of the emergent $U(1)$ symmetry as $\tilde{g}=0$ is an infrared stable fixed point.
So we have arrived at a conclusion that for TQCPs with time reversal global symmetry, strong coupling fixed points for gapless majorana fermions at TQCPs appear to have an emergent $U(1)$ symmetry
as far as interaction operators are local four-fermion ones. In $2d$, this turns out to have very important implications on the emergence of supersymmetry at these TQCPs.

Indeed, one can further find, due to such an emergent $U(1)$ symmetry, that $H'$ in this representation is identical to the Hamiltonian of two-component interacting Dirac fermions expressed in a majorana representation.
We therefore conclude that majorana fermions for TQCPs with time reversal symmetry in Eq.(\ref{TQCPTRI}), are equivalent to, up to a unitary transformation, conventional two component complex massless Dirac fermions with local interactions. 
They shall be completely equivalent in dynamics because the partition functions of two models are identical.  More explicitly, with time reversal symmetry at TQCPs, under the transformation defined in Eq.(\ref{UT}), (\ref{SUSY2}),

\begin{eqnarray}
&& H \rightarrow H'( \{ U(\frac{2\pi}{3}) \chi \}) =H_{2cD}(\{ \chi\}),
Z_{TQCP} =Z_{2cD}; \nonumber \\
&& H_{2cD} =\int d{\bf r}  [ \psi^\dagger({\bf r})  {\bf \sigma} \cdot i\nabla \psi ({\bf r}) +g_0 (\psi^\dagger\psi)^2], \psi^T= (\psi_\uparrow,\psi_\downarrow) \nonumber \\
\label{2cD}
\end{eqnarray}
where the subscribe $2cD$ refers two-component Dirac fermions, $Z_{TQCP}, Z_{2cD}$ are partition functions for TQCPs and 2-component Dirac fermions respectively.

Eq.(\ref{2cD}) appears to be surprising. TQCPs under considerations spontaneously break the original underlying U(1) symmetry of complex fermions and charges are not conserved because of superconductivity.
Gapless Dirac fermions on the other hand
do not break the U(1) symmetry and fermion charges are conserved in a Dirac model. 
The $U(1)$ symmetry discussed above is an emergent one only appear in the long wavelength limit. Nevertheless, its infrared physics of interacting gapless majorana fermions
appears to be identical to complex Dirac fermions because of this emergent symmetry. In $2d$, Dirac fermions with attractive interactions are supersymmetric at their strongly interacting fixed point.
If the emergent $U(1)$ symmetry above indeed also appears at a corresponding strong coupling fixed point, we intend to conclude
$2d$ TQCPs with time reversal symmetry where majorana fermions interact locally are also supersymmetric at its strong coupling fixed points.

In the limit we are interested in, all other local terms of $\chi^6, \chi^8$ vanish identically.
Any additional local interaction operators have to have a quartic form of $\chi^4$ form further involving gradient operators.  But they are irrelevant in the infrared limit from the renormalization point of view,
So generally if we restrict ourself only to local interactions, $U(1)$ symmetry appears to emerge in the infrared limit.

However, such an emergent $U(1)$ symmetry only appears when interactions can be completely expressed in terms of local operators, an assumption that might not be generic.
It is more natural that microscopic interactions actually break such a low energy symmetry at certain intermediate energy scales beyond the effective theories of local interactions.
In practical systems, interactions can be further mediated by various dynamic fields.
If we take this point of view, then although effective theories with local interactions
suggest an emergent $U(1)$ symmetry, actual strong coupling fixed point solutions generically shall not have such a $U(1)$ symmetry. 

This is especially so because low energy symmetries at gapless strong coupling fixed points
shall also be consistent with high energy symmetries/asymmetries of dynamic fields away from the strong coupling fixed points, in addition to low energy emergent symmetries.  
And as the fixed points are approached, if some of finite energy symmetries or asymmetries inherited by dynamic fields become descending in the infrared limit of TQCPs, 
they can results in breaking emergent $U(1)$ symmetries seen in the effective local theories above.
Below we are going to clarify this issue using a more general dynamic model.  
To understand details near strong coupling fixed point $\tilde{g}^*$ in $2d$ and $3d$, it is again more convenient to apply a generalized model introduced in the next subsection
which captures both the infrared physics discussed so far and additional higher intermediate energy sectors fully consistent with the infrared physics. These intermediate energy sectors eventually descend to low energy windows near 
strong coupling fixed points and dictate the behaviours of TQCPs in strong coupling limits.

\subsection{Time reversal symmetry invariant case: General}

To better understand the origin and consequence of emergent $U(1)$ symmetry and limitation of the above analyses,
we can further explore the TQCP dynamics in an extended majorana fermion model by explicitly introducing two real scalar fields $\phi_{1,2}=\phi_{1,2}^\dagger$,
\begin{eqnarray}
H &=& H_0 +H_I \nonumber \\
H_0 &=& \int d{\bf r}
[ \chi^T ({\bf r}) (\mathbb{I}  \otimes (\sigma_x  { i\nabla_x}+\sigma_z {i\nabla_z}) +\tau_y \otimes \sigma_y  {i\nabla y}) \chi ({\bf r}) 
 \nonumber \\
&+&\sum_{i=1}^{2}\pi^2_i({\bf r}) + \nabla \phi_i ({\bf r}) \cdot \nabla \phi_i({\bf r})+ M^2_i\phi_i^2({\bf r})]; \nonumber \\
H_I &=& \int d{\bf r}[ \sum_{i=1}^{2}g_{Yi} \phi_i({\bf r}) \chi^T({\bf r}) \tau_i \otimes \sigma_y  \chi({\bf r}) +g_4 (\sum_{i=1,2}\phi_i^2({\bf r}))^2 ]. \nonumber \\
\label{SUSY3}
\end{eqnarray}
where $\tau_{1,2}=\tau_{x,z}$, $M_{1,2}$ are masses of real scalar fields $\phi_{1,2}$ respectively.
And we have further set the speeds of real scalar fields, $v_{1,2}=1$ so to have a desired emergent Lorentz symmetry.
Two real scalar fields $\phi_{1,2}$ and four-component majorana fields in Eq.(\ref{SUSY3}) are defined in a standard way,

\begin{eqnarray}
& &[\phi_i({\bf r}), \pi_j({\bf r}')] =i\delta_{ij} \delta({\bf r}-{\bf r'}), \nonumber \\
&& [\phi_i({\bf r}),\phi_j({\bf r'}]=[\pi_i({\bf r}),\pi_j({\bf r}')]=0, i=1,2 \nonumber \\
& &\chi^T= (\chi_{1\uparrow}, \chi_{1\downarrow}, \chi_{2\uparrow},\chi_{2\downarrow}).
\end{eqnarray}

The Hamiltonian that describes TQCPs with time reversal symmetry is invariant under time reversal transformation ${\mathcal T}$.
In addition, just like an emergent $U(1)$ symmetry in Eq.(\ref{SUSY2}) for TQCPs with local interactions, it is also further invariant under an internal reflection transformation ${\mathcal R}$.
The reflection symmetry only emerges exactly at TQCPs and it is not a symmetry of gapped phases adjacent to TQCPs which only has ${\mathcal T}$-symmetry as we discussed in the previous section.
So at TQCPs, Eq.(\ref{SUSY3}) is invariant under
a bigger symmetry group of $Z_2 \otimes Z_2$.
 
Under the time reversal transformation ${\mathcal T}$, these real fields transform accordingly,

\begin{eqnarray}
\chi_1 & \rightarrow&  i\sigma_y K  \chi_1, \chi_2 \rightarrow -i\sigma_y  K \chi_2, \nonumber \\
 \phi_1 & \rightarrow& \phi_1,\phi_2 \rightarrow -\phi_2,  \pi_1 \rightarrow  \pi_1,\pi_2 \rightarrow -\pi_2
 \label{TRS}
\end{eqnarray}
where $K$ is an action of complex conjugate.
The Hamiltonian in Eq.(\ref{SUSY3}) is manifestly time reversal symmetric.

\begin{figure}
\includegraphics[width=8cm]{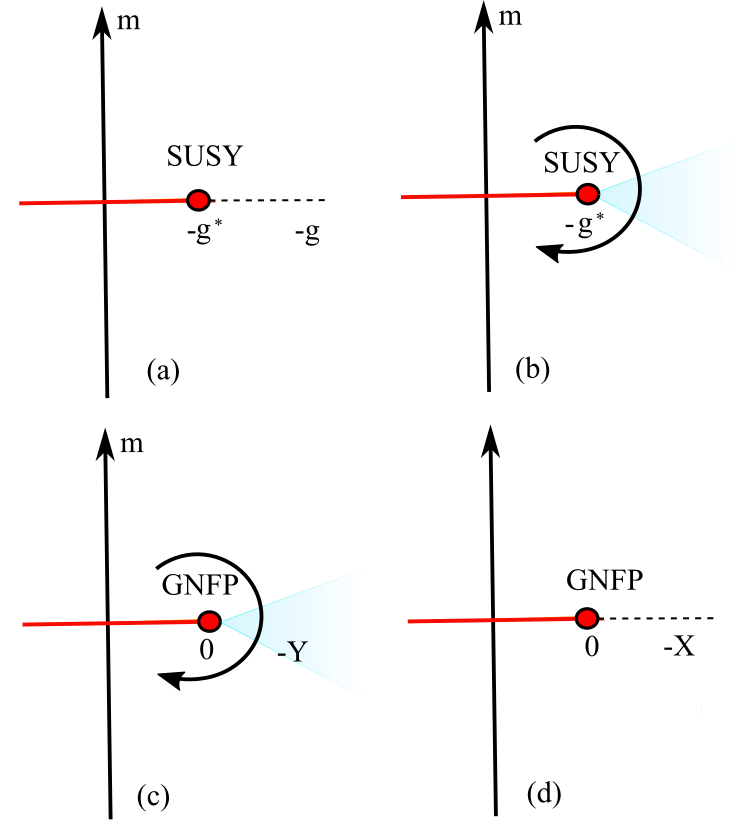}
\centering
\caption{ \footnotesize  $2d$ TQCPs in strong coupling limits in the $m-g$ plane ( (a) and (b) ), $m-Y$ plane (with $X=X_0 >0$) (c) and $m-X$ plane (with $ Y=Y_0>0$)  (d) . $m$ is the tuning fermion mass of TQCPs and $g$ is an effective local interaction constant in Eq.(\ref{SUSY1}),(\ref{SUSY2}). $X=M_1^2$, $Y=M_2^2$ are masses of scalar fields in dynamic models in Eq.(\ref{SUSY3}).
a) represents the results for $2d$ TQCPs wiithout time reversal symmetries; b-d) describe $2d$ TQCPs with time reversal symmetry. 
b-c) are scenarioes where time reversal symmetry is spontaneously broken in strong coupling limit beyond the red termination points.
d) is for TQCPs with strong interactions that do not result in spontaneously time reversal symmetry breaking.
In a)- d), the thick (red) line represents a $(d+1)$th transition lines obtained in Ref.\cite{Yang19} and dashed ones for first order transition lines.
The red dot  separating them in a) is supersymmetric (SUSY) with $N_f=\frac{1}{4}$ for TQCPs breaking time reversal symmetries but in d) is a non-supersymmetric {\em Gross-Neveu} fixed point (GNFP) . 
In b), the red termination point is supersymmetric with $N_f=\frac{1}{2}$ where majorana fermions interact with an emergent $U(1)$ symmetry; beyond the termination point, $U(1)$ symmetry is broken spontaneously.
In b),c), we also show possible paths around strong coupling fixed points that connect two topologically distinct phases without closing the fermion mass gap because of spontaneously breaking protecting time reversal symmetry beyond this point.
(Time reversal symmetry is broken spontaneously in shaded areas.) }
\label{SUSY}
\end{figure}

In addition, Eq.(\ref{SUSY3}) is also invariant under the following reflection transformation ${\mathcal R}$.

\begin{eqnarray}
\chi_1 &\rightarrow& - \chi_2, \chi_2 \rightarrow \chi_1 \nonumber \\
\phi_{1} &\rightarrow & -\phi_1,   \pi_1 \rightarrow  -\pi_1, \phi_2 \rightarrow -\phi_2,   \pi_2 \rightarrow -\pi_2
\label{RS}
\end{eqnarray}

In both $2d$ and $3d$,
Eq.(\ref{SUSY3}) admits four phases depending on the masses of scalar fields, $X=M_1^2$ and $Y=M_2^2$ (see Fig.\ref{U(1)}) and spontaneous symmetry breaking of $Z_2\otimes Z_2$ symmetries in  Eq.(\ref{TRS}),(\ref{RS}).\\
\\
i) When $X >0, Y>0$,  $\langle  \phi_1 \rangle=\langle \phi_2 \rangle=0$ and the phase is $Z_2 \otimes Z_2$ symmetric. Majorana fermions remain gapless; $X, Y \rightarrow \infty$ further corresponds to the weakly interacting limit studied in Ref.\cite{Yang19};
\\ \\
ii) When $X<0, Y > 0$, $\langle \phi_1 \rangle \neq 0$ but $\langle \phi_2 \rangle =0$ and the phase breaks the reflection ${\mathcal R}$ symmetry  but time reversal  ${\mathcal T}$ symmetry remains unbroken.
Symmetry group $Z_2 \otimes Z_2$ breaks down to $Z_2$ of ${\mathcal T}$ symmetry.
Majorana fermions acquire a time reversal invariant mass, $g_{Y1} \langle \phi_1 \rangle$ because of ${\mathcal R}$ symmetry breaking. This is an analogue of the well-know Gross-Neveu mechanism\cite{Gross74};
\\ \\
iii) When $X>0, Y<0$, $\langle \phi_2 \rangle \neq 0$ but $\langle \phi_1 \rangle =0$ and the phase again breaks $Z_2 \otimes Z_2$ symmetry to $Z_2$ of ${\mathcal RT}$.
Although both $Z_2$ of time reversal symmetry and $Z_2$ of reflection symmetry are spontaneously broken, this phase of matter is still invariant under the combined transformation ${\mathcal RT}$ of time reversal ${\mathcal T}$ and 
reflection ${\mathcal R}$. Majorana fermions acquire a time reversal symmetry breaking mass, $g_{Y2} \langle \phi_2 \rangle$.
\\ \\
iv) When $X<0, Y<0$, $\langle \phi_2 \rangle \neq 0$, $\langle \phi_1 \rangle \neq 0$. The phase spontaneously breaks $Z_2 \otimes Z_2$ symmetry and no symmetries remain.
Majorana fermions acquire two masses breaking both ${\mathcal T}$ and ${\mathcal R}$ symmetries.
Apparently, only iii) and iv) break the time reversal ${\mathcal T}$ -symmetry.
 \\ \\

The upper critical dimension of Eq.(\ref{SUSY3}) is $3d$, below which free majorana fermion fixed points are unstable when $M^2_1$ and/or $M^2_2$ becomes zero.
So in $2d$, Eq.(\ref{SUSY3}) has the following strong coupling fixed lines or points of majorana fermions when

\begin{eqnarray}
 &a)& \mbox{$X=0 \neq Y >0$ or   $M_1=0\neq M_2$ }\nonumber  \\
 &b)&\mbox{$Y=0 \neq X >0 $ or $M_2=0\neq M_1$ } \nonumber \\
 &c)& \mbox{$X=Y=0$, $M_1=M_2=0$.}
 \label{fps}
 \end{eqnarray} 
All these lines or points are time reversal symmetric and reflection symmetric and therefore are $Z_2 \otimes Z_2$ symmetric gapless states. In $2d$,
$M_1=0$ while $M_2 \neq 0$ or $M_2=0$ while $M_1 \neq 0$ represent conformal field theory states of Gross-Neveu fixed points (GNFP) in the model and we call them {\em GNFP}s or lines.
They correspond to $X=0, Y>0$ or $Y=0, X>0$ lines in Fig.\ref{U(1)} and form boundaries of gapless weakly interacting majorana fermions.

One can further verify that beyond the lines of fixed points a) or b), there is mass generation as in the Gross-Neveau (GN) model\cite{Gross74,Fei16}.
Beyond the line of fixed points b) or fixed point c), time reversal symmetry breaking masses are generated
while beyond the line of points a), only time reversal symmetric masses are generated. Along the line of $X=0, Y < 0$ and $Y=0, X <0$, majorana fermions remain massive;
$X=0, Y<0$ is ${\mathcal RT}$ $Z_2$ symmetric and $Y=0, X<0$ is $Z_2$ time reversal symmetric. 

In $3d$, phases displayed by Eq.(\ref{SUSY3}) are the same as in $2d$. However, the gapless fixed lines or points identified in Eq.(\ref{fps}) appear to be weakly interacting. In $3d$, $g_{Yi}$, $i=1,2$ actually flow into non-interacting fixed points
with $g_{Y1}=g_{Y2}=0$ instead of {\em CFT} fixed points. 

Eq.(\ref{SUSY3}) generally does not have a $U(1)$ symmetry because of an asymmetry between two masses $M_{1,2}$.
It can have a similar emergent $U(1)$ symmetry under a rotation around $\tau_y$ only if $M_{1,2}$ two masses are fully symmetric and $g_{Y1}=g_{Y2}$. 
For the purpose of our discussions below, without loss of generality, let us focus on the limit of $g_{Y1}=g_{Y2}$ at the moment although this condition can be relaxed later on when it comes to supersymmetry.

To further this discussion, we introduce the following $U(1)$ transformation

\begin{eqnarray}
\chi'=e^{i\tau_y \frac{\theta}{2}} \chi, \phi =e^{i\theta} \phi
\end{eqnarray}
where $\phi=\phi_1+i\phi_2$ is a complex field representation for scalar fields. 
In the gapped limit  when both $M_1 > M_2$ are nonzero, the long wavelength physics is identical to, as anticipated, Eq.(\ref{SUSY2}) after integration of massive scalar fields;
$H$ in Eq.(\ref{SUSY3}) again has an emergent $U(1)$ symmetry. 
However, in an intermediated scale between these two mass scales (assuming $M_1 > M_2$), $U(1)$ symmetry is always explicitly broken.
So for general masses $M_{1,2}$, $U(1)$ symmetry is absent in Eq.(\ref{SUSY3}),  unless $M_1=M_2$ (and $g_{Y1}=g_{Y2}$).

Only in the equal mass limit, $U(1)$ symmetry is present both in the infrared limit of $\Lambda < M_1=M_2$ and ultraviolet limit $\Lambda > M_1=M_2$.
The $U(1)$ symmetry above the mass gap is unique to the case of $M_1=M_2$ and is crucial for the symmetry in the gapless limit. It is this {\em ultraviolet symmetry of gapped scalar fields} that eventually 
emerge in the infrared limit as the gaps become lower and close and the state becomes gapless.

Consequently, such a $U(1)$ symmetry in general is absent in the gapless limit when only one of the masses, $M_1$ or $M_2$ becomes zero but the other one is non-zero, i.e. at fixed points a) and b) above. 
In the gapless limit, one can again verify that $H$ in Eq.(\ref{SUSY3}) is invariant under such a unitary rotation only when $M^2_1=M^2_2 \rightarrow 0$ as in case c).
In this limit, $g_{Y1} \rightarrow g_{Y2}$ as renormalization effects set in in the infrared scale  and don't require fine tuning. 
Furthermore, the strong coupling fixed point in this limit with a $U(1)$ symmetry shall also be supersymmetric as pointed out in previous studies\cite{Lee07,Fei16}
where strong evidence has been presented for $2d$. In other words, in $2d$.
among all fixed points a), b), and c) listed in Eq.(\ref{fps}) below Eq.(\ref{SUSY3}), only the supersymmetry fixed point c) has an emergent $U(1)$ symmetry.  
Later, we will apply these results to studies on TQCPs. 

So in $2d$, we intend to identify the fixed point in Eq.(\ref{RGE1}) that always has an emergent $U(1)$ symmetry, with the supersymmetry fixed point in the more general model of Eq.(\ref{SUSY3}) where full $U(1)$ symmetry only appears when $M_1=M_2$. Especially, when $M^2_1=M^2_2=0$ or at $X=Y=0$ in Fig.\ref{U(1)}, {\em CFT} states are a stable phase with an emergent supersymmetry of $N_f=\frac{1}{2}$ for general $g_{Y1},g_{Y2}$;
it effectively describes a four-component majorana fermion field or a two-component Dirac field interacting with a complex boson field $\Phi=\phi_1+i\phi_2$.

To summarize, we have observed that when a $U(1)$ symmetry emerges in a non-local dynamic model, strong coupling fixed points of gapless majorana fermion fields shall have, in addition to standard conformal symmetries, a supersymmetry.
At time reversal symmetric TQCPs defined in Eq.(\ref{TQCPTRI}), we can say that supersymmetry naturally emerges in strongly interacting majorana fermions  as a result of an emergent U(1) symmetry.
 
In more generic cases when a) $M_1 \neq M_2=0$ or b) $M_1=0\neq M_2$, an asymmetry in masses forbids an emergent $U(1)$ symmetry in the gapless limit.
The strongly interacting gapless majorana fermions in a) and b) shall just have standard scale-conformal symmetries as in {\em GNFP}s without additional higher symmetries.
$U(1)$ symmetries only emerge in the low wavelength limit when both masses $M_{1,2}$ are finite but do not appear when one of masses vanishes but the other remains finite.  
Therefore, {\em GN} fixed points in this limit shall be identified as TQCPs with time reversal symmetry where interactions between majorana fermions can not be simply characterized by local 
four-fermion operators.  In other words, when non-local interactions like the ones in Eq.(\ref{SUSY3}) are present to break the emergent $U(1)$ symmetry displayed in Eq.(\ref{TQCPTRI}), the corresponding 
dynamics in the gapless limit shall be in a non-supersymmetric {\em GNFP} class.

\begin{figure}
\includegraphics[width=8cm]{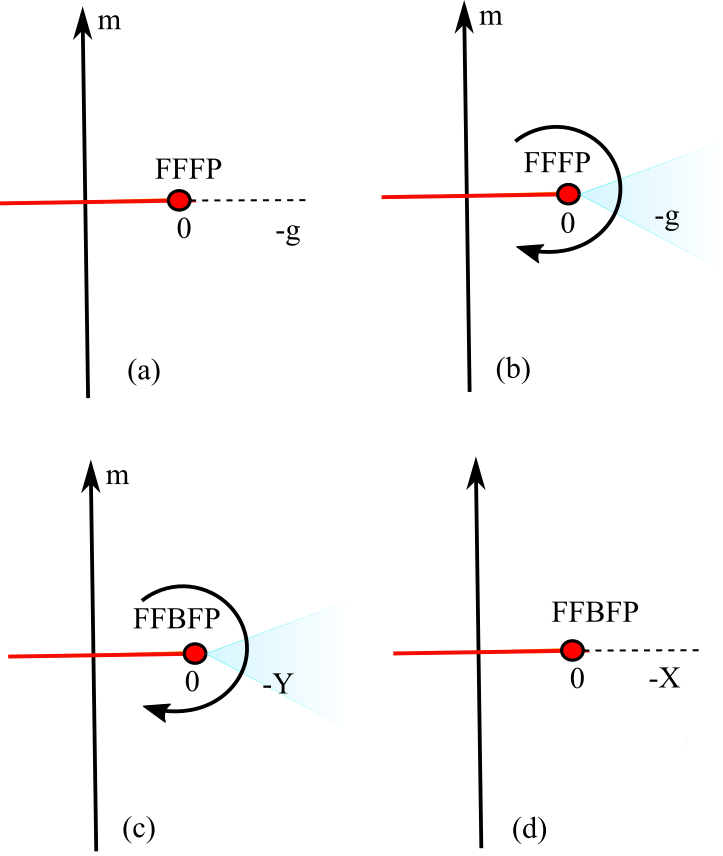}
\centering
\caption{ \footnotesize   $1d$ (a),b)) and $3d$ (c), d)) time reversal symmetric TQCPs in strong coupling limits. $m$ is the tuning fermion mass of TQCPs 
and $g$ is an effective interaction constant in models with local interactions, Eq.(\ref{SUSY3}),(\ref{SUSY2}), and $X=M_1^2$, $Y=M_2^2$ are masses of scalar fields in dynamic models in Eq.(\ref{SUSY3}).
a) is for $1d$ TQCPs when interactions result in spontaneous mass generation but do not result in time reversal symmetry breaking (i.e. symmetry breaking forced to be exactly along $\phi_1$  direction in Eq.(\ref{SUSY3})).
The thick (red) line represents a $(d+1)$th transition lines obtained in Ref.\cite{Yang19} and dashed ones for discontinuous first order transition lines.
b) represents $1d$ TQCPs with local interactions (i.e. Eq.(\ref{SUSY})) where time reversal symmetry is spontaneously broken in strong coupling limits beyond the red termination point (because of breaking of an emergent
$U(1)$ symmetry). In $1d$, all the termination points in a) and b) are free fermion fixed points (FFFP) unlike in $2d$.
In c)(with $X=X_0 >0$) and d)(with $Y=Y_0>0$) for $3d$, termination points are free fermion-boson fixed points (FFBFP) where new free gapless real bosons emerge in the infrared limit but decoupled from the free gapless majorana fermions. Paths in b) and c) connect two topological distinct states without closing the fermion mass gap because of spontaneous breaking of protecting symmetry; i.e., two states can effectively be smoothly deformed into each other.
(Time reversal symmetry is broken spontaneously in shaded areas.)
}
\label{SUSY3d}
\end{figure}

\section{Time reversal symmetric TQCPs in Strong Coupling Limits}

In our previous studies\cite{Yang19,Yang21}, we have identified that a broad class of TQCPs belong to weakly interaction majorana fermion classes, with or without Lorentz symmetry.
Now we are in a position to discuss those time reversal symmetric TQCPs in strong coupling limits and what happens to the free majorana fermion universality class. We will first focus on $2d$ which is the most interesting limit and come back to $1d$ and $3d$ later. Discussions in this section are entirely based on the results in Sec. VI on various strong coupling fixed points of majorana fermions.

\subsection{ $2d$ TQCPs in Strong Coupling Limits}

In case a) in the previous section, we drive TQCPs from a weakly interacting regime ($X>0, Y>0$ in Fig.\ref{U(1)} ) into a strongly interacting regime by varying $X=M^2_1$ across zero from a positive side $X=M^2_1 >0$ while holding $Y=M_2$ to be 
a constant.
Strong interactions in $X<0$ side result in a finite fermion mass, $g_{Y1} \langle \phi_1 \rangle$, that respects the time reversal symmetry. The ground state is gapped and two-fold degenerate because of spontaneously
$Z_2$ reflection symmetry breaking, and furthermore they are time reversal symmetric. 
The $(d+1)$th transition line previously pointed out in the weakly interacting limit\cite{Yang19} is therefore terminated by a standard {\em GN} fixed point  of type a)
beyond which transitions are first order and discontinuous. 

However, in case b) we drive TQCPs into strongly interacting limit by varying $Y=M^2_2$ from a positive side across zero while holding $X=M^2_1$ to be finite.
when $-\infty < g^2_{Y2}/M^2_2  < 0$,  time-reversal-symmetry breaking mass of fermions, $g_{Y2} \langle \phi_2 \rangle$, is spontaneously generated which is qualitatively different from case a) where this symmetry is respected. 
Strong interactions therefore can not only open a finite mass gap of fermions at TQCPs 
but further lead to a connection between two otherwise symmetry protected topological distinct phases via a symmetry breaking mass. 
Topological charges without protecting symmetries are ill defined and
one anticipates that two topologically distinct superconducting states can be effectively deformed into each other without closing the fermion mass gap in this limit\cite{LP}.
The $(d+1)$th transition line is therefore simply terminated at {\em GNFP}s of type b) in Eq.(\ref{fps})  and no {\em direct} transitions can occur beyond that point.

In c), we drive TQCPs from the weakly interacting regime into strongly interacting regimes by varying $M^2_1$ across zero while holding $M^2_1/M^2_2$ to be exactly one.
Time reversal symmetry breaking mass is generated as a result of spontaneously breaking of $U(1)$ symmetry when $\langle \phi_1\rangle^2 + \langle\phi_2\rangle^2$ remains constant.
This mass aspect is similar to case b). So the $(d+1)$th transition line is  again simply terminated at fixed point type c). However because of an emergent $U(1)$ symmetry,
the termination point is now supersymmetric with $N_f=\frac{1}{2}$. These results are summarized in Fig.\ref{SUSY}. 
Continuous weakly interacting time reversal symmetric TQCPs in $d=2$ therefore can be terminated by universal strong coupling fixed points a), b), c), beyond which there will be either no direct transitions if the protecting symmetry is broken spontaneously or discontinuous first order transitions.
Details further depend on dynamics and breaking of protecting time-reversal symmetry, and whether there is an emergent supersymmetry.
 
 \subsection{$1d$ TQCPs}
 
 So far, we have focused on time reversal symmetric TQCPs in $2d$. In $1d$, there is only a free fermion fixed point in Eq.(\ref{TQCPTRI}),(\ref{RGE1}) with $\tilde{g}=0$.
Local interactions are marginal as suggested in Eq.(\ref{TQCPTRI}), (\ref{RGE1}) and in this limit, there are no other strong coupling fixed points as in a standard upper critical dimension. Especially, interactions are marginally irrelevant when $\tilde{g} >0$ and marginally relevant when $\tilde{g} <0$. 
 This indicates that $\tilde{g}$ flows into $\tilde{g}^*=-\infty$ when $\tilde{g}$ is negative and 
 masses are spontaneously generated even at weak coupling limit through the well-known Gross-Neveu mechanism\cite{Gross74}. 
 On the other hand, free gapless majoarna fermions are stable as far as $\tilde{g}>0$.
   
Therefore for time reversal symmetric TQCPs (spinful case only) in $1d$,
$(d+1)$th order or $2$nd order transition lines in $1d$ only exist when $\tilde{g}>0$ and are terminated exactly at a free fermion fixed point $\tilde{g}=0$. Beyond the termination point, there shall be first order transitions if the protecting symmetry is unbroken
spontaneously. However, if interactions are strictly local, emergent $U(1)$ symmetry is broken beyond point $\tilde{g}=0$ and then there will be no direct transitions between two topologically distinct superconductors when $\tilde{g} <0$.
Instead, two topologically different states can again be smoothly deformed into each other without closing the fermion mass gap.
So the main difference between $1d$ and $2d$ is that termination points in $1d$ can not be {\em GNFP} or {\em SUSY} fixed points but are trivial free fermion fixed points ({\em FFFP}). Lines of $2$nd order TQCPs are now terminated by free majorana fixed points beyond which there are either $1$st order transitions or no {\em direct} transitions. At {\em FFFP}, dynamics are set by weakly interacting gapless majorana fermions.

\subsection {$3d$ TQCPs in Strong Couling Limits}

Just like in $2d$, there are strong coupling fixed points at $\tilde{g}^*=-(d-1)$ in $3d$ in addition to the free fermion fixed point. So again, strong coupling physics can be better understood in terms of emergent scalar fields Eq.(\ref{SUSY3}).
Eq.(\ref{SUSY3}) has the same phases as in $2d$ as shown in Fig.\ref{U(1)}.
However, it turns out in $3d$ that in the massless limit of Eq.(\ref{SUSY3}) when $X=0$ or $Y=0$ or $X=Y=0$, the Yukawa coupling constants $g_{Yi}$, $i=1,2$ flow into non-interacting fixed points instead of {\em GNFPs} or {\em SUSY} fixed points and the theory in this limit becomes trivial.
Effectively, in $3d$ at strong coupling fixed points $\tilde{g}^*=-2$,
there are well defined emergent gapless {\em real scalars} fields but completely decoupled from real fermions.
Strong interactions between majorana fermions near $\tilde{g}^*$ in $3d$ therefore mainly result in a new degree of freedom represented by free gapless real bosons.
The termination points of $(d+1)$th order continuous transition line in $3d$ 
 are now described by decoupled {\em gapless} free majorana fermions and {\em gapless} free real bosons and we call it a free fermion-boson fixed point ({\em FFBFP}).
 The gapless real bosons represent new emergent degrees of freedom, unique to strongly coupled majorana fermions in $3d$.
So all the termination points (red dots in Fig. \ref{SUSY} ) discussed in $2d$ are still present but instead of being conformal {\em GNFP}, they become
infrared trivial {\em FFBFP}. This concludes our discussions on $3d$ TQCPs with time reversal symmetries. And because of emergent real bosons, these termination points are also different from $1d$ termination points which 
are  represented by simple free gapless real fermions, without emergent gapless real bosons.
Results in $1d$ and $3d$ are summarized in Fig.\ref{SUSY3d}.

\section{Open Questions on Topological Quantum Criticality with Larger Global Symmetry Groups}

So far we have concentrated on TQCPs that either break all symmetries in a superconductor or with a (minimal) global symmetry of time reversal.
We find that TQCPs that break all global symmetries can be fully described by majorana fermion fields of $N_f=\frac{1}{4}$ while TQCPs with time reversal symmetry are characterized by majorana fields of $N_f=\frac{1}{2}$. 
Strong coupling fixed points of these gapless majorana fermions terminate
continuous  $(d+1)$th order phase transitions between topologically distinct states. 
Beyond these strong coupling fixed points, for TQCPs without time reversal symmetry, mass gap opens up signifying first-order phase transitions separated from the weakly interacting $(d+1)$th order transitions by supersymmetric fixed points.

For time reversal symmetric TQCPs, if time reversal symmetry respecting fermion mass gaps are opened up in strong coupling limits, ground states at TQCPs are expected to be time reversal symmetric and are two fold degenerate with $m= \pm |m|$. We again identify {these states} with first order transitions separating topologically distinct phases. A {\em CFT} state of {\em GNFP} class separates continuous $(d+1)$th order transitions line in the weakly interacting limit from the discontinuous transition line.

We have also found that for time reversal symmetric TQCPs, time reversal symmetry breaking majorana mass gaps can also be spontaneously generated in the limit of strong coupling.
In $2d$, the $(d+1)$th order transition line is then completely terminated by strong coupling {\em Gross-Neveu} fixed points beyond which states become simply connected without closing gaps. 
If majorana fermions at TQCPs only interact locally, additional emergent $U(1)$ symmetry leads to supersymmetric termination points. Spontaneous $U(1)$ symmetry breaking simultaneously leads to breaking of protecting time reversal symmetry of topological superconductors and topological states can again be deformed smoothly.

What happens to superconducting TQCPs in $d=2,3$ with larger global symmetry groups remains to be systematically studied in the future. When symmetry groups get larger, degrees of freedom associated with majorana fields also increase 
linearly leading to power-law large number of {\em local} interaction operators. These operators can form different representations of subgroups of global symmetries and dimensions of interaction parameter space become
much higher than one dimension or two dimensions that we have so far focused on in this article. 
In that case, there are no fundamental principles that forbid Green's functions discussed in Sec. II from developing more peculiar structures {\em without spontaneously breaking protecting symmetries}. For instance, the bulk-(gapless) boundary correspondence 
in weakly interacting limits that relies on an index theorem can be severely violated because of emergence of zeros in Green's functions in strongly interacting limits\cite{Gurarie11}.
So an interacting topological state with protecting symmetries might be deformed into ones with different topological invariants without a transition.

 As TQCPs appear to be infrared stable in $d=2,3$ where local fermion interaction operators are infrared irrelevant, the bulk-gapless boundary correspondence near weakly interacting TQCPs shall remain valid.
If smooth deformation between states in $d=2,3$ with different topological invariants does happen {\em while protecting symmetry is not spontaneously broken}, one can speculate that it likely involves some strong coupling fixed points where the standard bulk-gapless boundary relation starts breaking down.
Under which specific conditions of interactions and when this can happen or can not happen in a topological superconductor need to be investigated in the future.
Occurrence of smooth deformation in strong coupling limit also suggests that the topological-invariant approach outlined in Sec. II in general is inadequate and  
one has to introduce more general topological classification to properly represent strongly interacting topologically matter with certain global symmetries\cite{Fidkowski10,Fidkowski11,Fidkowski13,Metlitski15,Wang15,Song17,Xu21}.

The possibility of such a scenairo can be closely related to the well-known Fidkowski-Kitaev path of {\em escape} in one spatial dimension\cite{Fidkowski10}. There a global time reversal symmetry is embedded in a much bigger ($so(8)$) symmetry group unlike what we have in this article. Using {\em eight} chains of majorana fermions, they have successfully constructed $so(7)$-subgroup invariant on-site interaction operators embedded in $so(8)$ representations and utilized them to determine an $so(7)$ symmetric line boundaries between first order transitions and no-transition regions, in addition to an $so(8)$ line boundary separating regions of second phase transitions and no-transitions. These highly symmetric line boundaries have been applied to explicitly identify a path that smoothly connects two gapped topologically different states in a parameter space. In addition, emergent gauge symmetries and gauge fields discussed in Ref.\cite{Bi19} can also be relevant to TQCPs in certain strong coupling limits although it remains to be investigated when they become so in topological superconductors and superfluids.

\section{Conclusions}

In conclusion, we have discussed the important role of global symmetries at topological quantum critical points (TQCPs). The global symmetries determine the numbers of emergent fermion fields at a generic TQCP.
The number corresponds to $N_f=\frac{1}{4}$ or a single two-component majorana field, when all symmetries including $U(1)$ global gauge symmetry are broken, and is $N_f=\frac{1}{2}$ or a four component majorana fermion field when a time reversal 
symmetry is present at a TQCP without additional global symmetries. This number will increase when more symmetries are present at TQCPs until it reaches a maximum number of $N_f=N/2$ where $N (=2n)$ is the number of microscopic complex fermions available at a TQCP.
In general, the number of emergent low energy majorana fermion fields at a TQCP is independent of the number of complex fermions in a solid, $N$,  and only depends on global symmetries. 
We have presented our results in both cases with or without time reversal symmetry. This is also expected to lead to distinctly different thermodynamics and dynamics in quantum critical regimes\cite{Yang21b}. 

We have also discussed strong coupling limits of TQCPs. We have found that without time reversal symmetry, $2d$ TQCPs as strong coupling fixed points 
are supersymmetric with $N_f=\frac{1}{4}$. Furthermore, the supersymmetric {\em CFT} fixed point terminates the weakly interacting $(d+1)$th order continuous transition line and separates it from a first order order transition line beyond the supersymmetric state. Two topologically different superconductors in this case are always separated by a transition line even in strong coupling limits.

However, there are two different classes of strong coupling {\em CFT} states for $2d$ TQCPs with one global symmetry of time reversal invariance.
One of them has an emergent $U(1)$ symmetry and is supersymmetric with $N_f=\frac{1}{2}$. They describe strongly coupling physics of majorana fermions only interacting locally near a time reversal symmetric TQCP.  
In general, time reversal symmetric strong coupling TQCPs are {\em CFT} states without higher symmetries such as 
supersymmetry or $U(1)$ symmetry; they correspond to {\em Gross-Neveu} fixed points ({\em GNFPs}) with $Z_2\times Z_2$ symmetries of time reversal ${\mathcal T}$ and reflection ${\mathcal R}$. 

In the limit we have focused on if time reversal symmetry is not broken spontaneously, two topological states are either separated by a $(d+1)$th continuous transition line or a first order phase transition
and can not be deformed into each other smoothly. The strong coupling fixed points that separate them are always {\em GNFPs}.
However, if protecting time reversal symmetry is spontaneously broken in the presence of strong interactions, then the continuous transition lines emitted from weakly interacting limits will be simply terminated beyond which states can be deformed 
into each other while the fermion mass gap remains open. Termination points of $(d+1)$th continuous transition lines in $2d$ can be either supersymmetric with $N_f=\frac{1}{2}$ or non-supersymmetric {\em GNFPs} depending on whether the time reversal symmetry breaking occurs via spontaneous $U(1)$ symmetry breaking or not. 

In $1d$, continuous transition lines of $(d+1)$th order time reversal symmetric TQCPs are always terminated by a free majorana fermion fixed point beyond which there is a first order phase transition if protecting 
symmetry is not broken spontaneously or no direct transitions if symmetry is broken.
In $3d$ time reversal symmetric TQCPs, these termination points, beyond which there are either first order transitions or no transitions again depending symmetry breaking, are strongly interacting and can be effectively characterized by
emergent free gapless real bosons weakly coupled to free gapless majorana fermions. 

To summarize, at TQCPs without any global symmetries, we find that two topologically distinct states are always separated by transitions even in strong coupling limits. At TQCPs only with time reversal symmetry but no other global symmetries, we find that two topologically distinct states are always separated by transitions even in strong coupling limits we have focus on if the protecting symmetry is not broken spontaneously.
Smooth deformation without gap closing becomes possible only when time reversal symmetry is broken spontaneously. 

The author would like to thank Ian Affleck, Zhengcheng Gu, Sung-Sik Lee and Xiao-Gang Wen for inspiring discussions and Fan Yang for discussions on regularizing interacting majorana fields.
He also wants to thank the TDLi workshop on "New Frontiers in Extremely Strongly Interacting Matter" in July, 2021 where interactions and topology were discussed
and the KITP workshop on Topological Matter: Atomic, Molecular and Optical Systems (topology21) for its hospitality.
This project is in part support by a Professional Development Grant from the University of British Columbia, and by an NSERC (Canada) Discovery grant under contract RGPIN-2020-07070.

\end{document}